\documentclass[10pt]{IEEEtran}
\usepackage{graphicx,subfig}
\usepackage[export]{adjustbox}
\usepackage{amssymb}
\usepackage{epstopdf}
\usepackage{listings}
\usepackage{courier}
\usepackage{color}

\usepackage{parskip}
\usepackage{url}
\let\labelindent\relax
\usepackage{enumitem}
\usepackage{datetime2}

\usepackage[utf8]{inputenc}

\title{Coronavirus Contact Tracing: Evaluating The Potential Of Using Bluetooth Received Signal Strength For Proximity Detection }
\author{Douglas J. Leith, Stephen Farrell\\
    School of Computer Science \& Statistics,\\ 
    Trinity College Dublin, Ireland\\
	6th May 2020
    }

\begin{document}
\maketitle


\begin{abstract}  
We report on measurements of Bluetooth Low Energy (LE) received signal strength taken on mobile handsets in a variety of common, real-world settings.  We note that a key difficulty is obtaining the ground truth as to when people are in close proximity to one another.   Knowledge of this ground truth is important for accurately evaluating the accuracy with which contact events are detected by Bluetooth LE.  We approach this by adopting a scenario-based approach.  In summary, we find that the Bluetooth LE received signal strength can vary substantially depending on the relative orientation of handsets, on absorption by the human body, reflection/absorption of radio signals in buildings and trains.  Indeed we observe that the received signal strength need not decrease with increasing distance.  This suggests that the development of accurate methods for proximity detection based on Bluetooth LE received signal strength is likely to be challenging.   
Our measurements also suggest that combining use of Bluetooth LE contact tracing apps with adoption of new social protocols may yield benefits but this requires further investigation.  
For example, placing phones on the table during meetings is likely to simplify proximity detection using received signal strength.  Similarly, carrying handbags with phones placed close to the outside surface.   
In locations where the complexity of signal propagation makes proximity detection using received signal strength problematic entry/exit from the location might instead be logged in an app by e.g. scanning a time-varying QR code or the like.

\end{abstract}

\section{Introduction}

There is currently a great deal of interest in the use of mobile apps to
facilitate Covid-19 contact tracing.  This is motivated by the hope that more efficient and scalable contact
tracing might allow the lockdown measures currently in place in many countries to be relaxed more quickly~\cite{ferretti2020quantifying}.   

In this report we take a first step in evaluating the potential for using Bluetooth Low Energy (LE) received signal strength to detect periods of close contact between people.   We present measurements taken on mobile handsets in a variety of common, real-world settings.  We also measure the effect on received signal strength caused by the human body, by different types of indoor wall, the relative orientation of mobile handsets and so on.

The basic idea of a contact tracing app is that if two people carrying mobile handsets installed with the app spend significant time in close proximity to one another (e.g. spending 15 minutes within 2 metres\footnote{Ideally, what counts as a contact event should be informed by the conditions under which the virus is actually spread e.g. perhaps there is less risk of infection when people are 1m apart but facing away rather than towards each other, or perhaps contact with surfaces is more important than transmission through the air.  However data to inform more refined definitions of a contact event currently seems to be lacking. }) then the apps on their handsets will both record this contact event.  If, subsequently, one of these people is diagnosed with Covid-19 then the contact events logged on that person's handset in the recent past, e.g. over the last two weeks, are used to identify people who have been in close contact with the infected person.  These people might then be made aware of the contact and advised to self-isolate or take other appropriate precautions.   For this approach to be effective it is, of course, necessary that the app can accurately detect contact events.   

Almost all modern handsets are equipped with Bluetooth LE wireless technology and there is currently much interest in using this as the means for detecting contact events.   The Singapore TraceTogether contact tracing app~\cite{bluetracewhitepaper,ot,tt} is perhaps the first widely used app that uses Bluetooth LE for detecting contact events.   More recently, Apple and Google have formed a partnership to develop contact event detection based on Bluetooth LE~\cite{applegoogle}.

TraceTogether uses the Bluetooth LE received signal strength to estimate proximity, and likely the Apple/Google API will do the same.   In general, a radio signal tends to get weaker as it gets further from the transmitter since the transmit power is spread over a greater area.   Bluetooth LE devices can be configured to transmit \emph{beacons} at regular intervals and the idea is that the signal strength with which a beacon is received provides a rough measure of the distance between transmitter and receiver.  Namely, when the received signal strength is sufficiently high then this may indicate a contact event and, conversely, when the received signal strength is sufficiently low then this may indicate that the handsets are not in close proximity.

However, the propagation of radio signals in practice is often complex, especially in indoor environments where walls, floors, ceiling, furniture etc can absorb/reflect radio waves and so change the received signal strength.   A person's body also absorbs Bluetooth LE radio signals so that the received signal strength can be substantially reduced if their body lies on the path between the transmitter and receiver.

With this in mind, we report here on measurements of Bluetooth LE received signal strength taken on mobile handsets in a variety of common, real-world settings.  A key difficulty in evaluating proximity detection accuracy in real-world settings is establishing ground truth i.e. recording when contact events actually happened.  This ground truth is needed so that the contact events flagged by a contact tracing app can be compared against the actual contact events and so allow the accuracy of the app at detecting contact events to be assessed.   To address this we generally adopt a scenario-based approach.  For example, we take measurements as two people walk a circuit in city streets side by side while maintaining a 1 metre distance between them.   This has the great advantage that (i) ground truth is clear (to within experimental error, e.g. people will not be able to maintain an exactly 1 metre distance while walking) and (ii) data is collected in real-world settings with all their associated complexity.   The disadvantage of course is that this limits our study to fairly simple, well structured scenarios.  However, by selecting scenarios that aim to capture some of the key elements in common activities we can still gain useful insight into the real-world performance of Bluetooth LE received signal strength for proximity detection.

We present measurements for four real-world scenarios: people walking outdoors in city streets, people sitting around a meeting table, people sitting in a train carriage and people grocery shopping in a supermarket.   In addition we present supporting measurements evaluating the impact of the relative orientation of handsets on received signal strength, and on the signal attenuation caused by the human body, by a woman's handbag and by different types of indoor wall.

In summary, we find that (i) the Bluetooth LE received signal strength can vary substantially depending on the relative orientation of handsets and on absorption by the human body.  This means that, for example, the received signal strength is considerably higher when people are walking side by side compared to when they walk one behind the other at the same distance.   Similarly, we find that when people are sitting around a meeting table with their phones in their trouser pockets the received signal strength is low even when the people are sitting less than 1m apart.   When they place their phones on the table the received signal strength is much increased.   We also find that (ii) within complex indoor environments the received signal strength need not decrease with distance and indeed may increase with distance (recall that we generally expect signal strength to decrease with increasing distance, not to increase).   For example, for two people walking around a large supermarket we find that the received signal strength is much the same when they walk close together and when they walk 2m apart.   We observe increases in signal strength with increasing distance within a domestic house and within a train carriage.   
On a more positive note we find that (iii) outdoors the received signal strength indeed tends to decrease with increasing distance thus facilitating proximity detection, also that blockwork walls strongly attenuate Bluetooth LE radio signals and so leakage of these signals between adjacent houses/buildings is likely to be small. 

While we consider a much wider range of real-world scenarios and our focus is on contact tracing, our observations are consistent with previous measurement studies on the use of Bluetooth for proximity detection for other purposes.  For example,~\cite{face2face2014,iccsac2015,percom2017,sensys2019} in the context of measuring social interaction and ~\cite{conext2017,conext2018,ipin2018} in the context of indoor localisation.   These previous studies have generally found simple thresholding of received signal strength to be highly error prone and therefore found it necessary to use machine learning methods trained on data collected from the particular environment of interest in order to obtain accuate proximity detection.  These methods depend upon the availability of ground truth data, and the ability of the developed methods to generalise beyond the specific setting considered is not clear.

This suggests that the development of accurate methods for proximity detection based on Bluetooth LE received signal strength is likely to be challenging.   Apps based on Bluetooth LE are therefore probably not a panacea but rather are best viewed as a potentially useful addition to existing contact tracing methods.   Our measurements also suggest that combining use of Bluetooth LE contact tracing apps with adoption of new social protocols\footnote{Such social protocols should, however, not be used to effectively make it mandatory for people to carry a handset with a contact tracing app installed, e.g. employers should generally not make this mandatory for all employees.} may yield benefits although this requires further investigation.  For example, placing phones on the table during meetings is likely to simplify proximity detection using received signal strength.  Similarly, carrying handbags with phones placed close to the outside surface. 
In locations where the complexity of signal propagation makes proximity
detection using received signal strength problematic, one might consider
logging entry to/exit from the area in an app.  This would allow the Bluetooth data collected in that area to be flagged as being less reliable, hopefully reducing the rate of contact tracing errors.  Logging of entry/edit might, for example, be achieved by scanning a time-varying QR code, tapping a handset on a near-field tag (similar to contactless payment) or by placing a  dedicated Bluetooth beaconing device at the entry/exit point whose beacons can be logged\footnote{Such entry/exit logging is compatible with existing decentralised contact tracing architectures.  Namely, a record of the codes associated with these entry/exit events plus the times when they occurred could be stored locally on a user's handset.  Upon a person being discovered to be infected their recorded codes could be uploaded and other people can check these against their own locally stored record.  The codes used should be changed fairly frequently to avoid linking of codes with locations.   Entry/exit logging may, however, require modification of app APIs}.   





\section{Brief Overview of Bluetooth LE}
Bluetooth Low Energy (LE) was standardised in 2010.  The low energy moniker refers to the reduced drain on the device battery compared to the older Bluetooth Classic technology. The first mobile handsets using Bluetooth LE appeared in 2011-12 (e.g. the iPhone 4S) and today almost all modern handsets come equipped with it.  

Bluetooth LE operates in the same 2.4GHz unlicensed radio band as WiFi and other devices (including microwave ovens).   Bluetooth LE devices advertise their presence by periodically (typically once per second) broadcasting short beacon messages.  To mitigate the effects of interference from other users of the 2.4GHz band each beacon is broadcast simultaneously on three widely spaced radio channels.   

Each beacon essentially consists of a short fixed \emph{preamble}, followed by a small beacon \emph{payload}.  The payload contains an identifier of the device making the boadcast (in modern devices this identifier is usually randomised and changes frequently to improve privacy) plus a short message (generally up to 31 bytes long).  This message is typically used to indicate that the beacon is associated with a particular app or service, e.g. to associate it with a contact tracing app.

A device equipped with a Bluetooth LE receiver scans the three beacon radio channels listening for beacon transmissions.   When the start of a transmission is detected the receiver uses the fact that the beacon preamble is fixed and known to fine tune the radio receiver to the incoming signal.  As part of this fine tuning process a received signal strength indicator (RSSI) is output, which is an estimate of the radio power in the received signal.   It is worth noting that this RSSI measurement is intrinsically noisy, with fluctuations of $\pm 5dB$ or greater common even in situations with simple line-of-sight radio transmission, e.g. see ~\cite{conext2017} and Figure \ref{fig:garden} below.  If the received signal strength is too weak either the transmission is simply not noticed or this fine-tuning process fails.  Typically this occurs when the received signal strength is below around -90dB (the noise floor of the receiver).   Upon successful fine-tuning of the receiver the payload of the beacon is decoded and passed up to the operating system and then on to relevant apps.

The received signal strength is affected by the transmit power used by the device broadcasting the beacon.   Bluetooth LE devices generally use a relatively low transmit power (to save on battery drain) and a rough guideline is that beacons cannot be decoded at distances beyond about 10 metres from the transmitter.  In practice the received signal strength is, however, also greatly affected by the way in which the radio signal propagates from transmitter to receiver.  In general the radio signal gets weaker as it travels further since the transmit power is spread over a greater area.  However, many complex effects can be superimposed upon this basic behaviour.  In particular, obstacles lying on the path between the transmitter and receiver (furniture, walls etc) can absorb and/or reflect the radio signal and cause it to be received with higher or lower signal strength.  A person's body also absorbs radio signals in the 2.4 GHz band and so the received signal strength can be substantially reduced if their body lies on the path between the transmitter and receiver.   In indoor enviroment walls, floors and ceilings can reflect radio signals even when they are not on the direct path between transmitter and receiver, and so increase or decrease the received signal strength.   

Metal, in particular, strongly reflects radio waves and this can be an important factor in radio propagation in environments with a lot of metal.  In buses and trains the walls, floor and ceiling are mainly metal and the seats often contain metal parts.  In supermarkets not only are the shelving, fridges and freezers typically made of metal but also tinned groceries etc located on the shelving.  We can therefore expect that radio propagation in these environments will be complex, and in particular due to reflections the signal strength may not decrease as quickly with distance as in other environments e.g. see~\cite{trains2009,trains2019}.

\section{Baseline Measurements}

\subsection{Hardware \& Software Used}
In our tests we used two pairs of mobile handsets: a pair of Google Pixel 2's and a pair 
of Samsung Galaxy A10's, both running Android 9.   Since we only acquired the Galaxy A10's after our measurement study had already started most of our measurements were made using the Pixel 2's, and unless otherwise stated the measurements that we report below are Pixel 2 data.

We used the OpenTrace app~\cite{ot} to collect measurements.  This is an open source version of the TraceTogether~\cite{tt} contact tracing app used in Singapore.   We slightly modified OpenTrace to (i) dump the measured receive signal strength values to disk in csv format and (ii) to scan/advertise Bluetooth beacons more frequently (allowing us to collect data more quickly, albeit at the cost of an increased drain on the handset battery which is why these settings would not be used in a production contact tracing app).   Note that in our initial experiments we used the default scan/advertise settings of OpenTrace, which yields a receives signal strength measurement roughly every 10s.  The change (ii) above increases the sample rate to a measurement roughly every second.  We verified that this caused no other change to the collected received signal strength values.

\subsection{Signal Attenuation With Distance}
To provide a baseline we collected measurements of received signal strength vs distance.   We placed two handsets at the same height and collected measurements of received signal strength as we varied the spacing between them.   At each distance the handsets are held in a fixed position for at least 10 minutes while the measurements are taken, giving around 80 signal strength observations (one for each advertised beacon detected) at each distance.  The handsets are placed flat with the screen facing upwards (as we will see later the relative orientation of the handsets can affect the received signal strength).

\begin{figure}
\centering
\subfloat[RSSI vs distance]{
\includegraphics[width=0.45\columnwidth,valign=t]{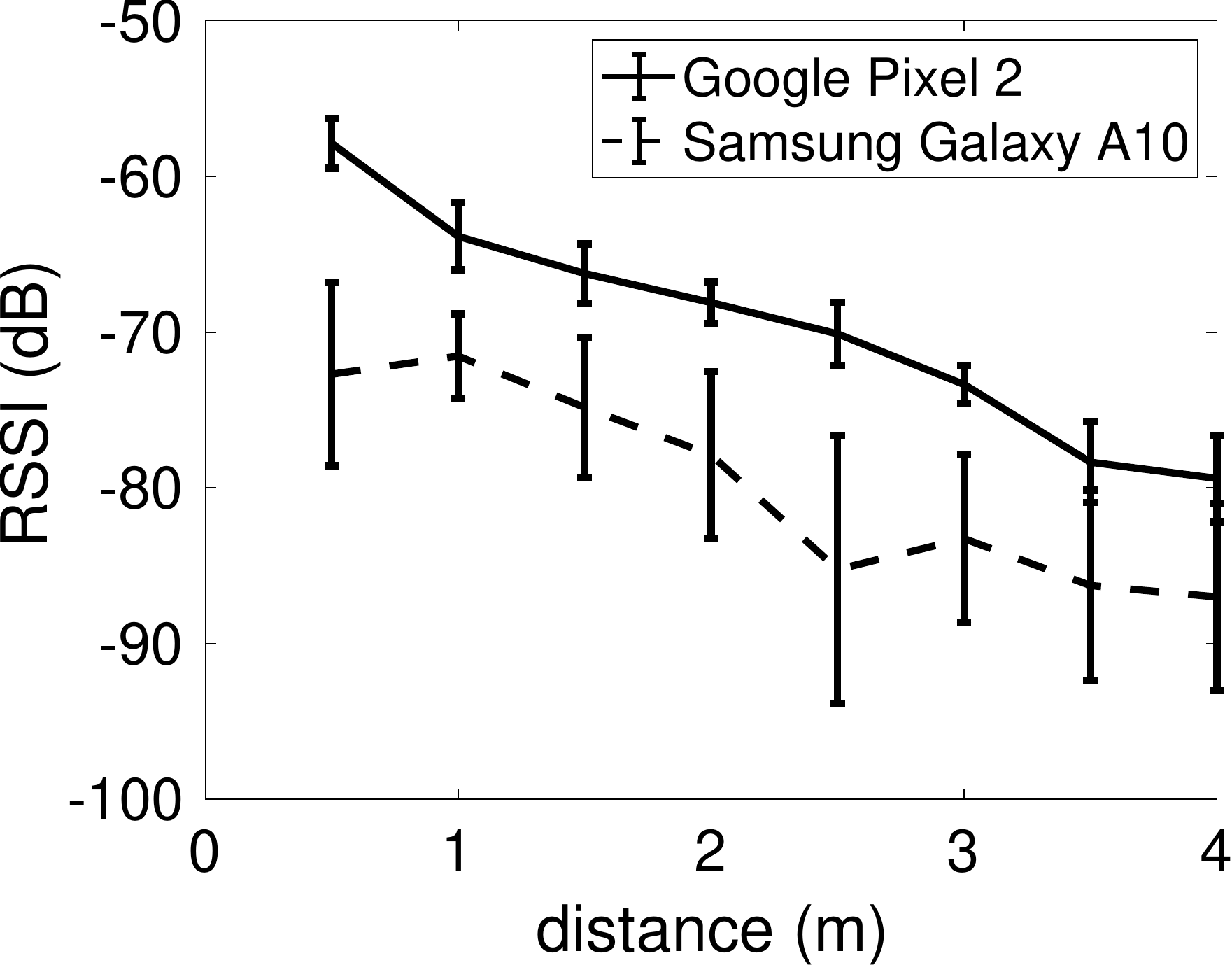}
}
\vspace{0.5cm}
\subfloat[Garden location]{
\includegraphics[width=0.39\columnwidth,valign=t]{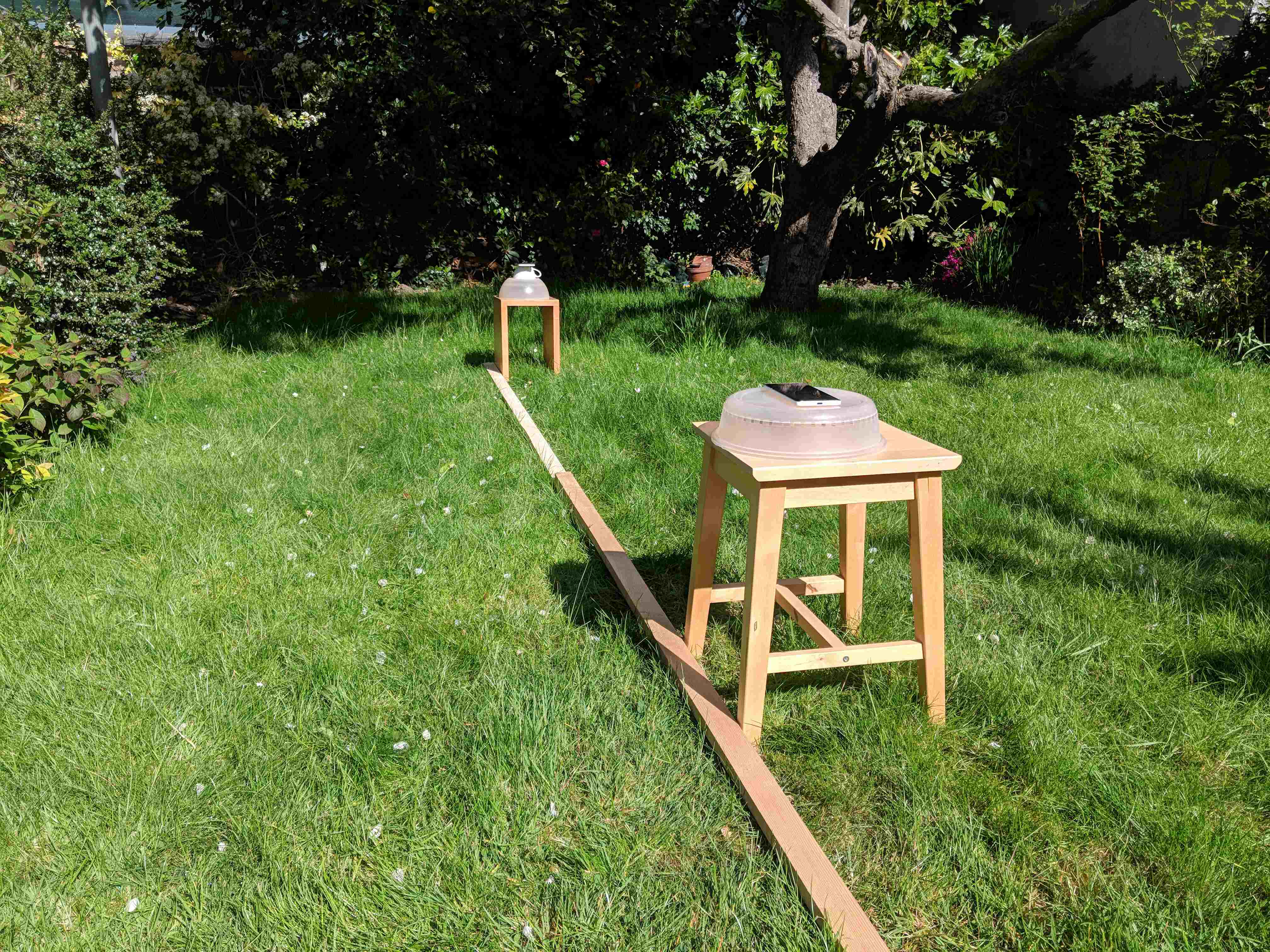}
}
    \caption{(a) Measured received signal strength (RSSI) vs distance.  Pixel 2 measurements were collected at the outdoor location shown in (b), the Galaxy A10 measurements were collected in a different but similar outdoor location.}\label{fig:garden}
\end{figure}


\begin{figure}
\centering
\subfloat[RSSI vs distance]{
\includegraphics[width=0.45\columnwidth,valign=t]{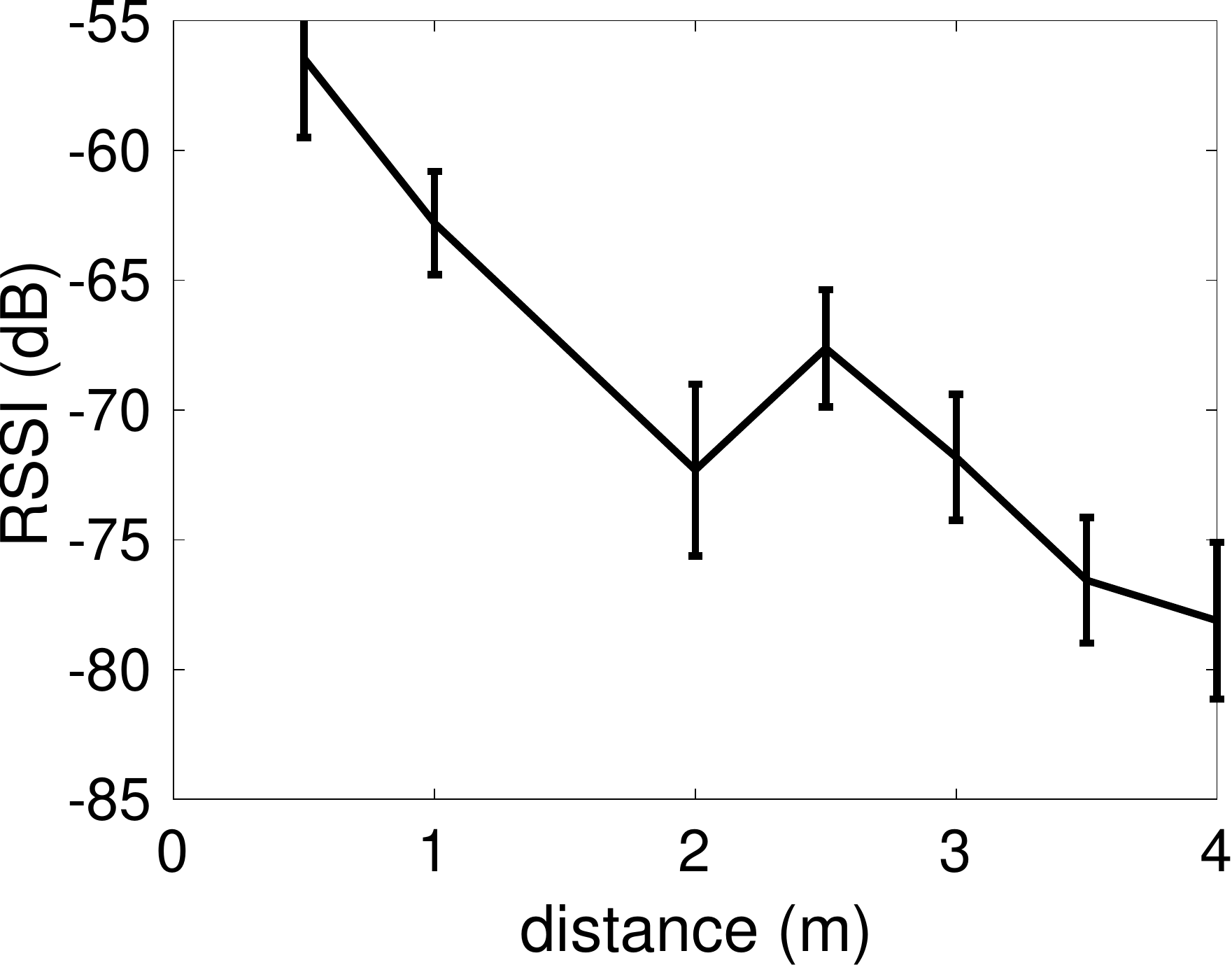}
}
\subfloat[Indoor location]{
\includegraphics[width=0.39\columnwidth,valign=t]{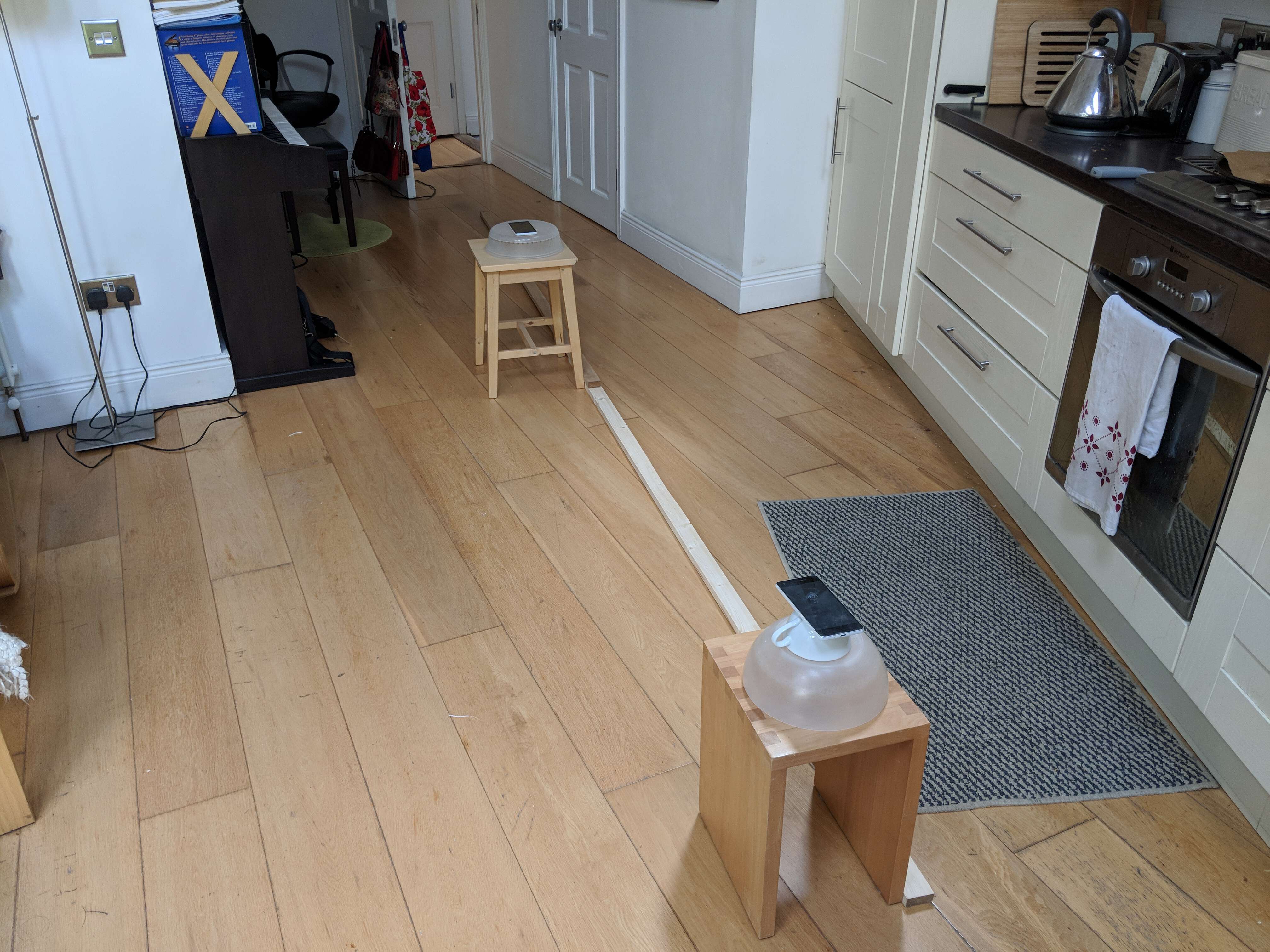}
}
\caption{(a) Measured received signal strength vs distance.  Measurements collected in a domestic indoor space shown in (b).  Note the narrowing of the space at the top of the photo.}\label{fig:kitchen}
\end{figure}


The solid line in Figure \ref{fig:garden}(a) shows the measured received signal strength vs distance for our pair of Google Pixel 2 handsets placed in an open space outdoors (the location is shown in Figure \ref{fig:garden}(b)).   It can be seen that the received signal strength decreases steadily with increasing distance, as expected.  Also shown are error bars indicating one standard deviation of the measured signal strength at each distance.  It can be seen that the standard deviation is around 5dB, consistent with previous observations in the literature, e.g. see~\cite{conext2017}. 

Since the received signal strength depends on the transmit power and antenna gain of the transmitter and also on the antenna gain at the receiver it can be expected that the measured received signal strength may vary with the handset model.  The dashed line in Figure \ref{fig:garden}(a) shows received signal strength vs distance for a pair of Samsung Galaxy A10 handsets.   Observe that the received signal strength is lower than with the Pixel 2's (presumably their transmit power and/or antenna gain differs) and also the standard deviation is somewhat higher at around 10dB.  We also collected measurements for a Huawei P10 and the RSSI vs distance curve (not shown in the figure to avoid clutter) is almost the same as that for the Pixel 2 but again with somewhat higher standard deviation.

It is also important to note that due to reflections from walls, furniture etc wireless signal propagation indoors is usually more complex than it is outdoors.    Figure \ref{fig:kitchen}(a) shows measurements of received signal strength vs distance taken in a relatively open domestic indoor space (shown in Figure \ref{fig:kitchen}(b)).  Observe the \emph{increase} in received signal strength in Figure \ref{fig:kitchen}(a) as the distance increases from 2 to 2.5m.  This effect is consistent and reproducible using multiple devices, it is not a measurement error.  We believe that it is associated with the narrowing of the indoor space that can be seen towards the top of Figure \ref{fig:kitchen}(b), with the walls (formed from concrete blockwork covered in plaster) acting to focus the wireless signal and so increase the received signal strength as the handset is moved from the open room into this narrower space.   We have also observed similar effects outdoors.   Such behaviour has obvious implications for the use of received signal strength to measure proximity, and confirms that caution is needed when interpreting received signal strength.

\subsection{Signal Attenuation By Human Body}
\begin{figure}
\centering
\subfloat[Received signal strength ]{
\includegraphics[width=0.5\columnwidth,valign=c]{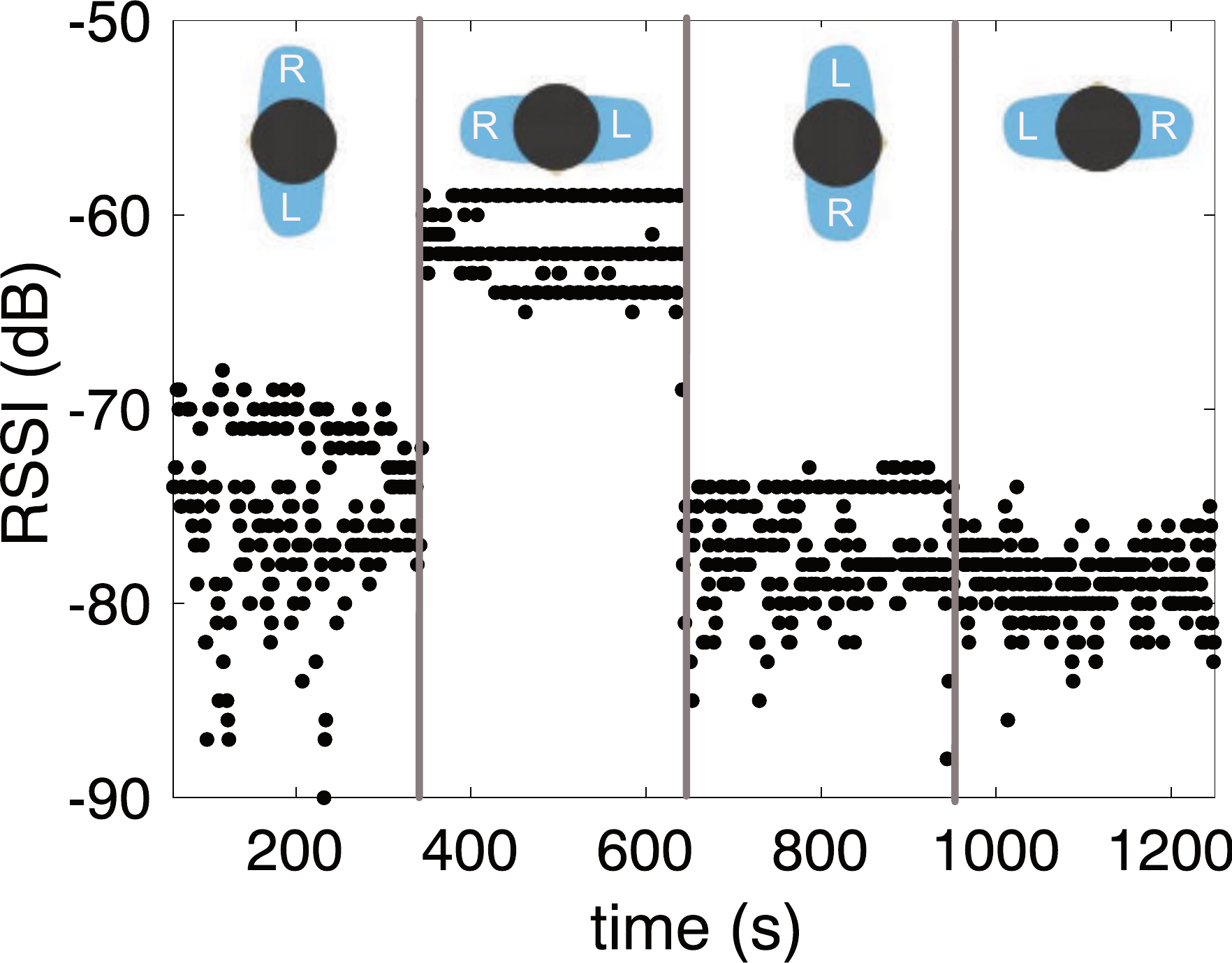}
}
\subfloat[Setup]{
\includegraphics[width=0.25\columnwidth,valign=c]{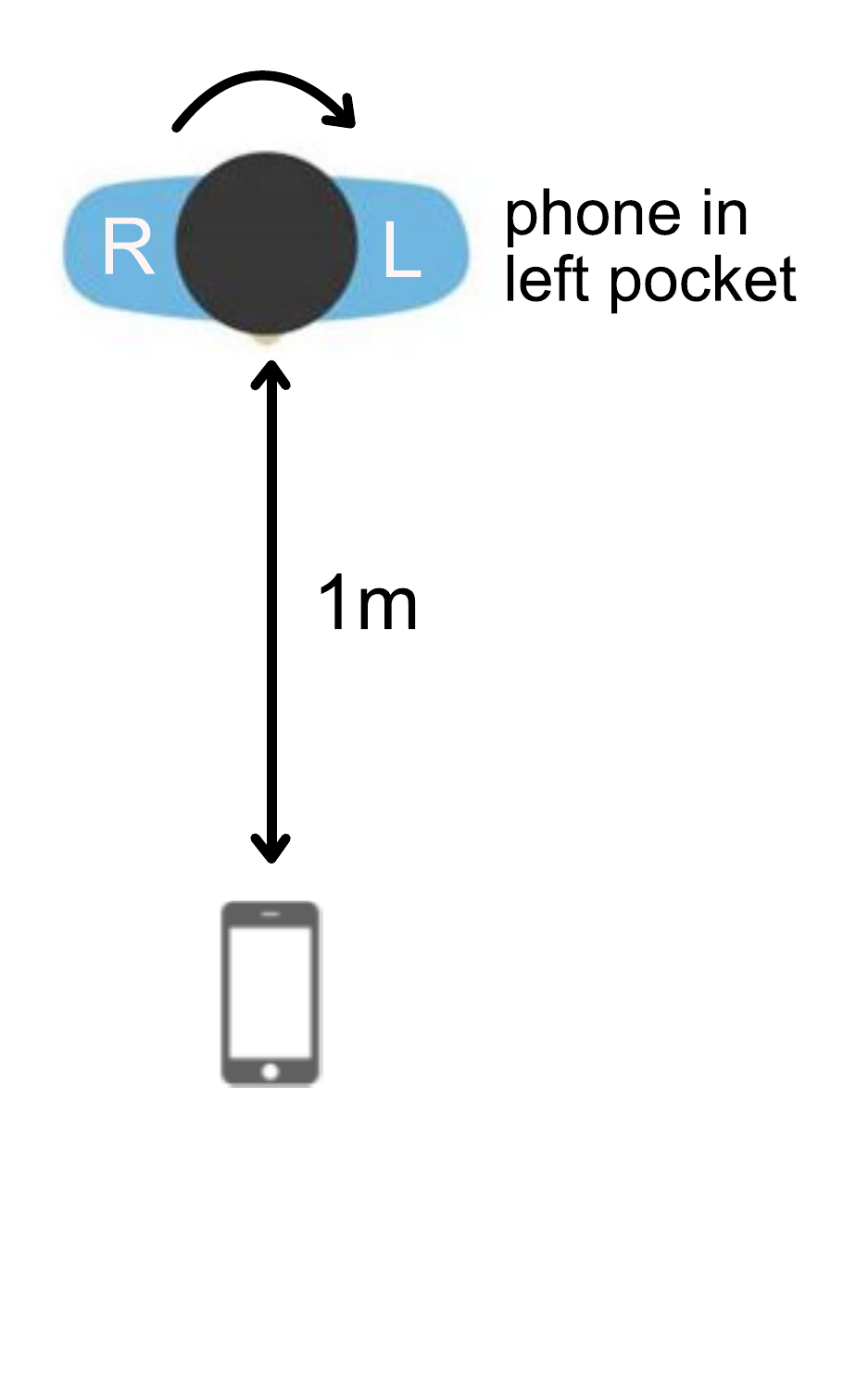}
}
\caption{Measured received signal strength as a person rotates and changes their orientation relative to a handset held fixed at 1m distance.}\label{fig:orient_pocket}
\end{figure}

We also expect that attenuation of Bluetooth LE wireless signals by the human body (Bluetooth LE transmits at 2.4GHz, a frequency which is absorbed by water molecules hence why it is also used in microwave ovens) may affect received signal strength, and that the relative orientation of handsets may matter.  Figure \ref{fig:orient_pocket}(a) shows measurements of received signal strength as a person rotates around a fixed point 1m away from a mobile handset places flat on a wooden table.  The person carries a second handset in their left trouser pocket.  Figure \ref{fig:orient_pocket}(b) shows the experimental setup schematically.   

It can be seen from Figure \ref{fig:orient_pocket}(a) that the received signal strength varies by around 20dB as the person rotates.    The received signal strength is slightly higher (around -75dB) when the person's left side is oriented towards the fixed handset than when their left-side is oriented away from the handset (around -80 dB), as might be expected since in the latter case the person's body lies between the phone in their left pocket and the fixed handset.   Observe also that the received signal strength is substantially higher (around -60dB) when the person is facing the fixed handset than when they have their back to it (around -80dB), again presumably due to signal absorption by the person's body (the pocket is located towards the front of their trousers).

\begin{figure}
\centering
\includegraphics[width=0.5\columnwidth]{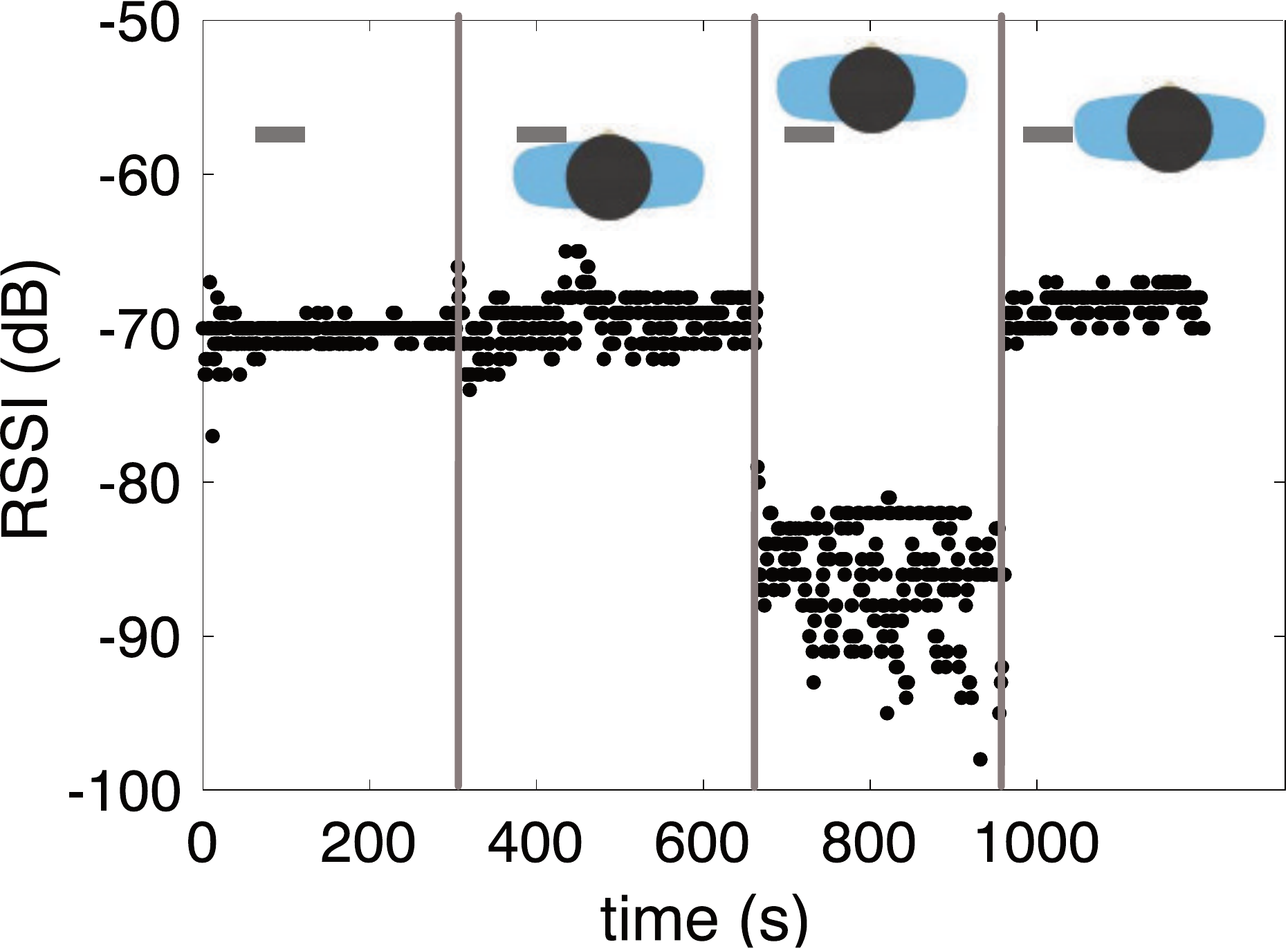}
\caption{Measured received signal strength between two handsets spaced 1m apart as a person changes position relative to the handsets (the positions of the handsets remains fixed).  In the left-hand section of the plot nobody is near the handsets, in the centre-left section a person stands closely behind one of the handsets (mimicking a handset being in a front trouser pocket), in the centre-right section they stand closely in front of the handset (mimicking a handset being in a rear trouser pocket) and in the right section they stand beside the handset.}\label{fig:human}
\end{figure}

In Figure \ref{fig:orient_pocket}(a) both the signal path between the two handsets and their relative orientations change.  To separate out these effects we also took measurements with two handsets held in fixed positions at a 1m distance and roughly waist height (1m above the ground).   Figure \ref{fig:human} shows the measured received signal strength as a person takes up various positions close to one of the handsets.   Perhaps unsurprisingly, it can be seen that when the person stands close in front of the handset (roughly mimicking a handset being in a rear trouser pocket) the received signal strength is around 15dB lower than when the person stands closely behind the handset (roughly mimicking a handset being in a front trouser pocket).   

We also took measurements with a person holding a phone in a fixed position and orientation at chest height at a 1m distance from a fixed handset.  The received signal strength with the person facing the fixed handset, so with an unobstructed path between the two handsets, was observed to be around 10dB higher than when the person faced away from the fixed handset so that their torso lay on the signal path.

\subsection{Signal Attenuation By Woman's Handbag}

\begin{figure}
\centering
\subfloat[Received signal strength ]{
\includegraphics[width=0.5\columnwidth]{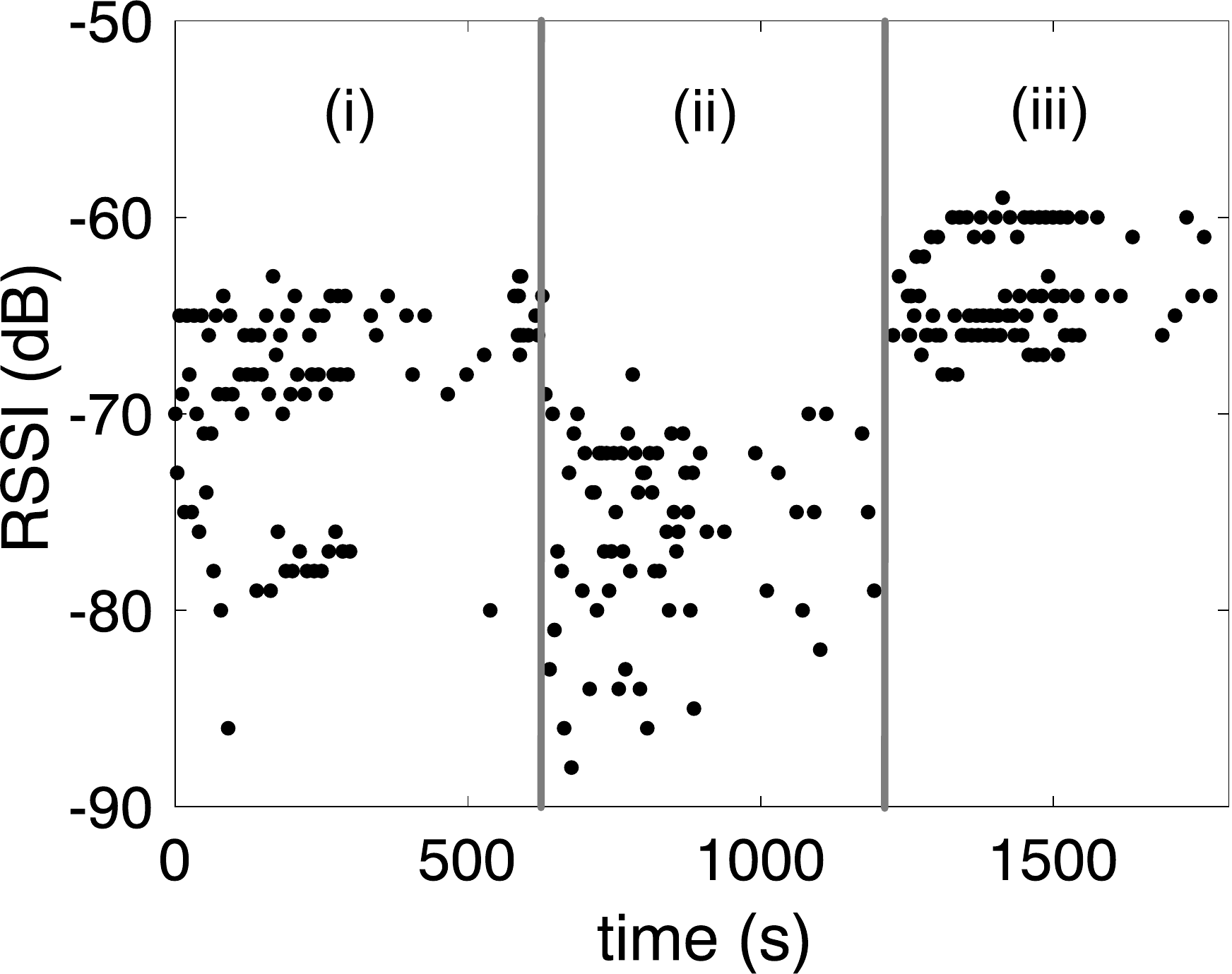}
}
\subfloat[Setup]{
\includegraphics[width=0.25\columnwidth]{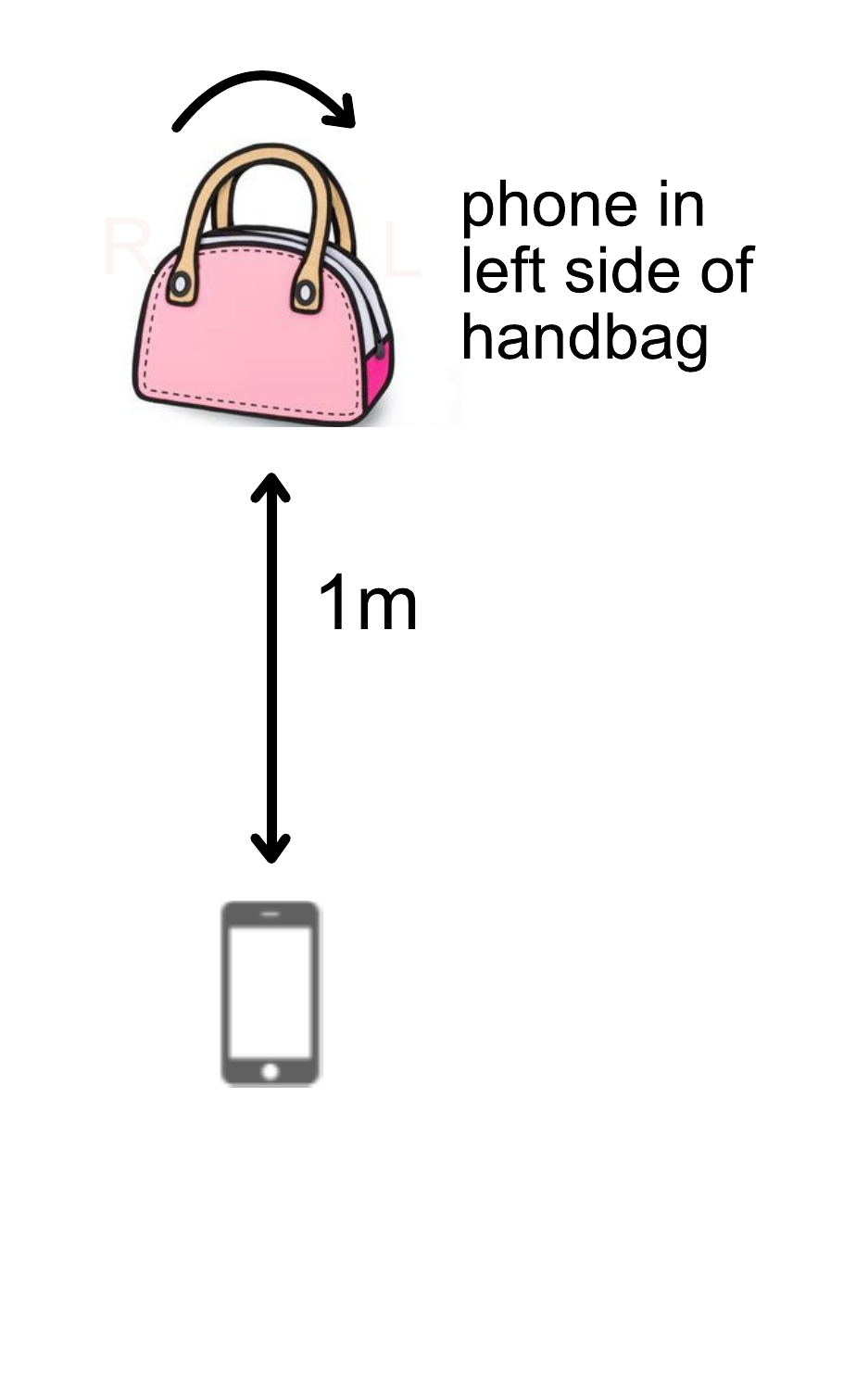}
}
\caption{Measured received signal strength between a phone placed within a cloth handbag and a second phone at a 1m distance.   The phone is located at the left-hand side of the handbag and measurements are collected with (i) the side of the bag containing the phone oriented towards the second phone, (ii)  the side of the bag containing the phone oriented away the other phone (i.e. with a 180$^\circ$ rotation of the bag from the initial position (i)) and (iii) with the handbag edge on to the second phone (i.e. a 90$^\circ$ rotation from the initial position (i)).}\label{fig:handbag}
\end{figure}

Figure \ref{fig:handbag}(a) plots received signal strength measurements for a similar setup to the previous section but now with one handset placed inside a handbag rather than in a trouser pocket.  The handbag contains other items (purse, cosmetics, hairbrush etc) and the phone is located within a pocket inside the left side of the bag.   A second phone is placed 1m from the handbag and the measurements are taken as the handbag is rotated, see Figure \ref{fig:handbag}(b) for a schematic of the setup.

In section (i) of Figure \ref{fig:handbag}(a) the handbag is orientated so that left-hand side is facing the second phone and a signal strength of around -68dB is observed.  The handbag is then rotated 180$^\circ$ so that the right-hand side of the bag now faces the second phone, section (ii) of the plot.  It can be seen that the signal strength falls by about 10dB to around -75dB.  The handbag is now rotated by 
90$^\circ$ so that it is end on to the second phone, section (iii) of the plot, and this change increases the signal strength to around -65dB.

\subsection{Signal Attenuation By Walls In A Building}

\begin{figure}
\centering
\subfloat[Stud partition]{
\includegraphics[width=0.45\columnwidth,valign=t]{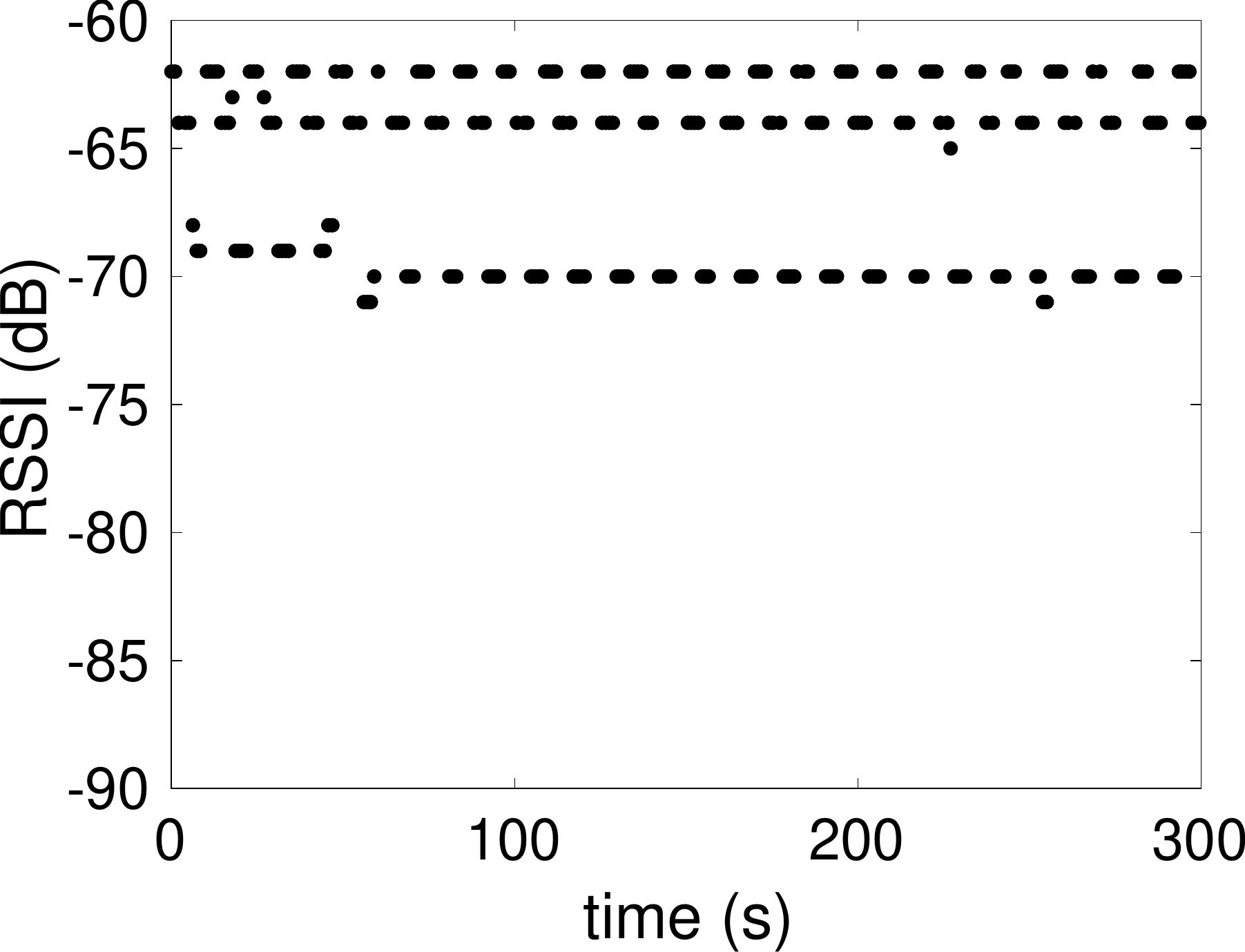}
}
\subfloat[Blockwork wall]{
\includegraphics[width=0.45\columnwidth,valign=t]{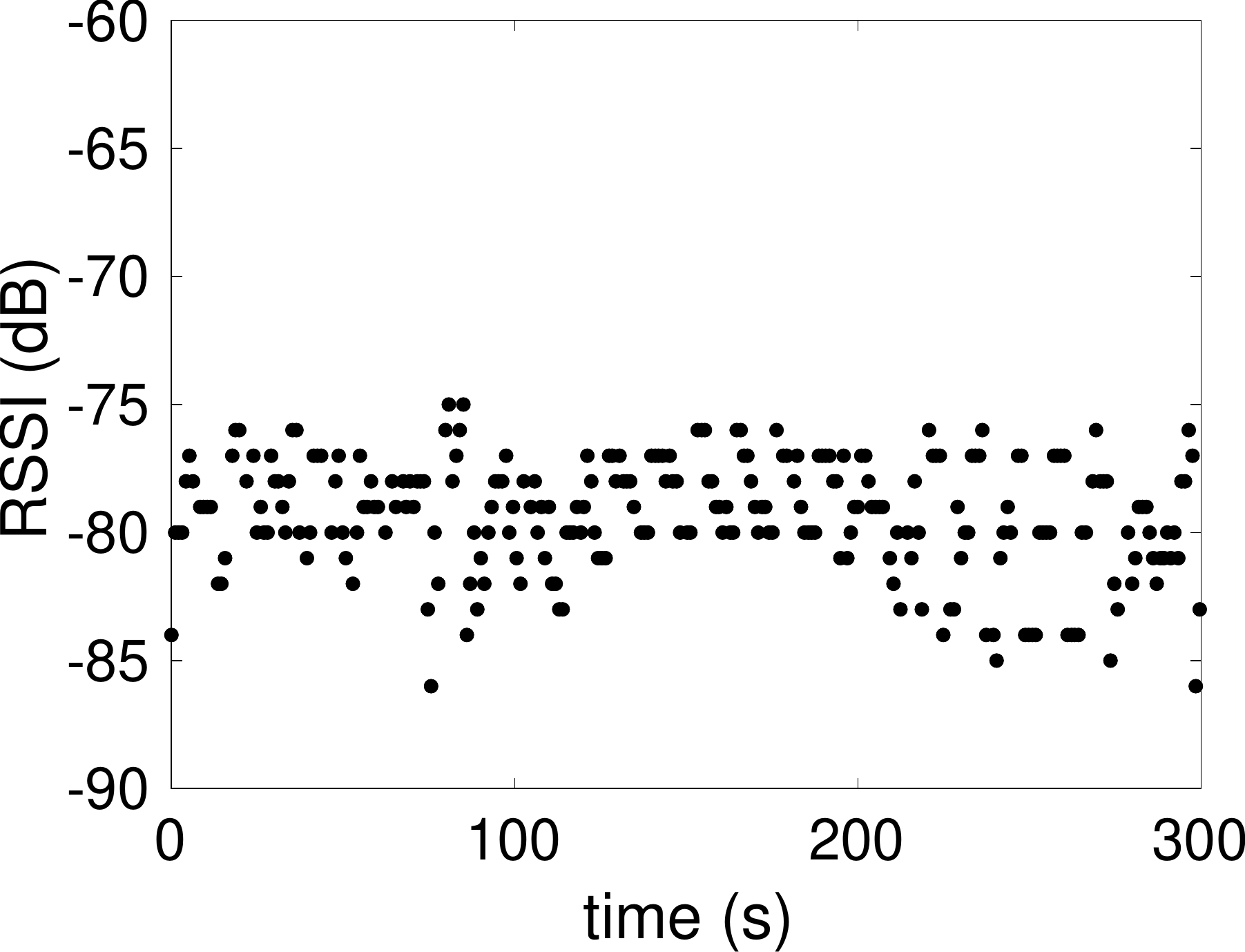}
}
\caption{Impact of two types of wall on measured received signal strength.  Handsets are placed approximately 1m apart with wall in between, (a) measurements for a wooden stud wall surfaced with plasterboard and (b) shows measurements for a blockwork wall approximately 15cm thick.  }\label{fig:walls}
\end{figure}

Figure \ref{fig:walls} illustrates the impact of walls on signal propagation.   Measurements are shown for a plasterboard stud wall and for a 15cm thick blockwork wall in a 1930s Dublin house.   Comparing Figure \ref{fig:walls}(a) with the 1m point in Figure \ref{fig:garden}(a) it can be seen that the stud wall creates little signal attenuation.  In contrast, it can be seen from  Figure \ref{fig:walls}(a) that the blockwork wall attenuates the signal by around 20dB.   Blockwork is commonly used for the party walls separating semi-detached houses and apartments.   These measurements therefore suggest there is little danger of Bluetooth signal strength data triggering a false alarm regarding proximity between people located in adjacent houses/apartments.  Stud walls, however, are widely used internally within buildings and our measurements indicate that they have little effect on Bluetooth received signal strength.  Hence, false alarms regarding proximity between people located in adjacent rooms within the same building may be a concern.



\section{Scenario-Based Measurements}
\subsection{Scenario 1: Walking In City Streets}
Our first scenario seeks to evaluate proximity measurement between people walking in city streets.  The observations from this scenario are probably also applicable to people walking in parks and large indoor spaces such as shopping centres and airports.   

We collected measurements of Bluetooth LE received signal strength for two people walking the same 1.5km circuit along suburban streets in Dublin in four different configurations: side by side (shoulders touching), side by side maintaining a 1m gap, one behind the other maintaining a 1m gap and a 2m gap.   Both people carry a mobile handset in their left-hand trouser pocket.  

\begin{figure}
\centering
\subfloat[Side by side]{
\includegraphics[width=0.45\columnwidth]{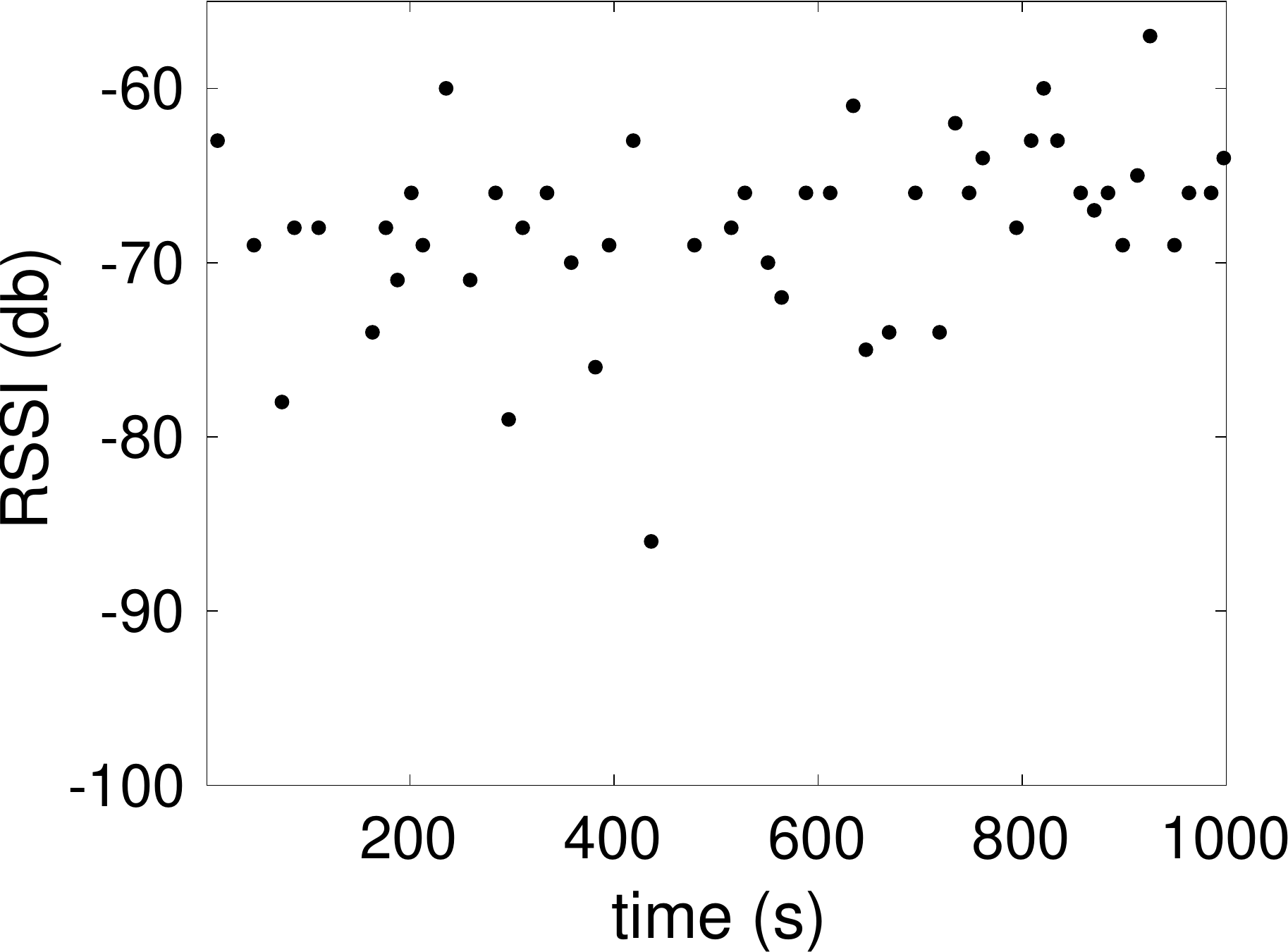}
}
\subfloat[1m to side]{
\includegraphics[width=0.45\columnwidth]{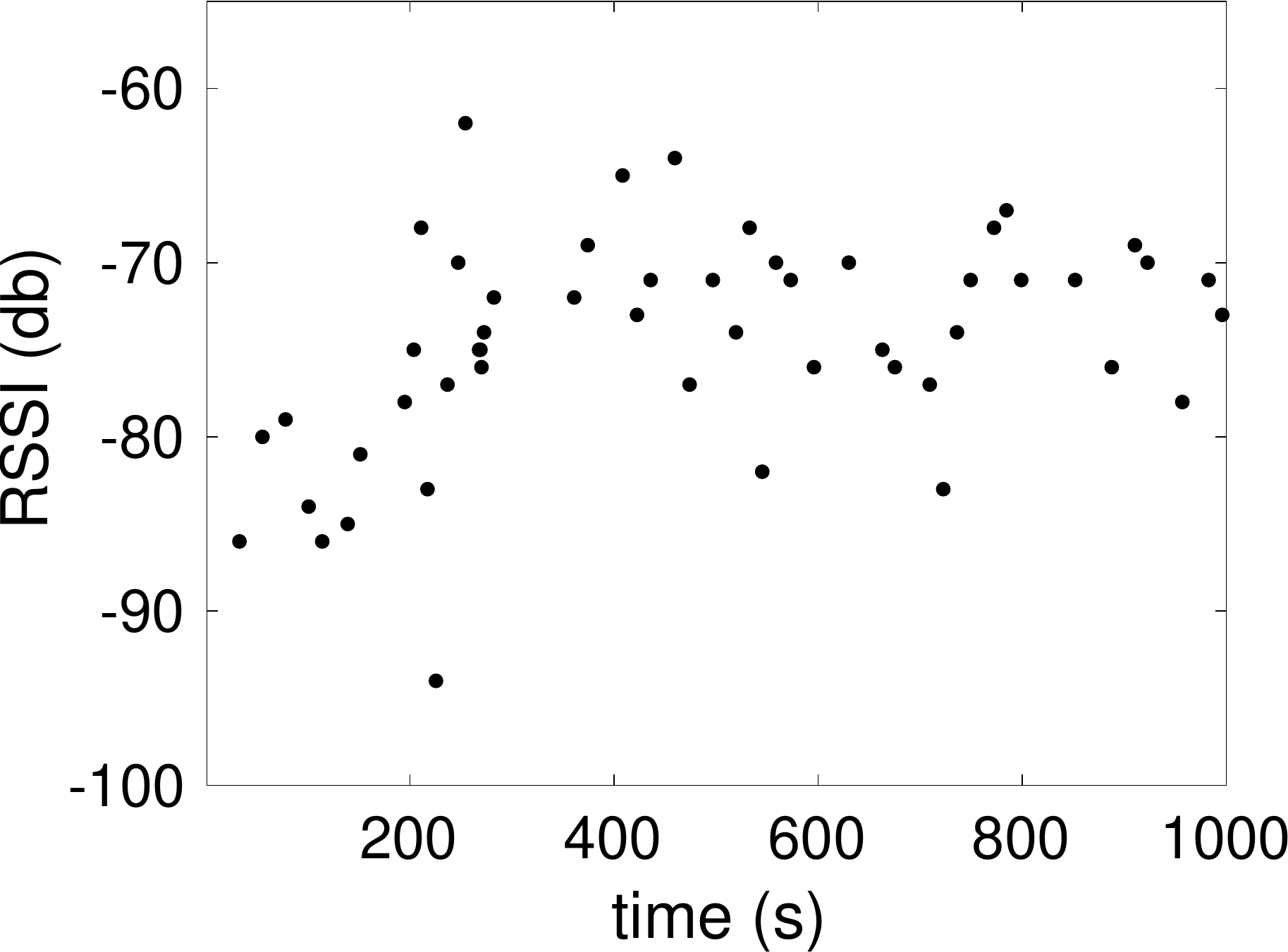}
}
\vspace{0.2cm}
\subfloat[1m behind]{
\includegraphics[width=0.45\columnwidth]{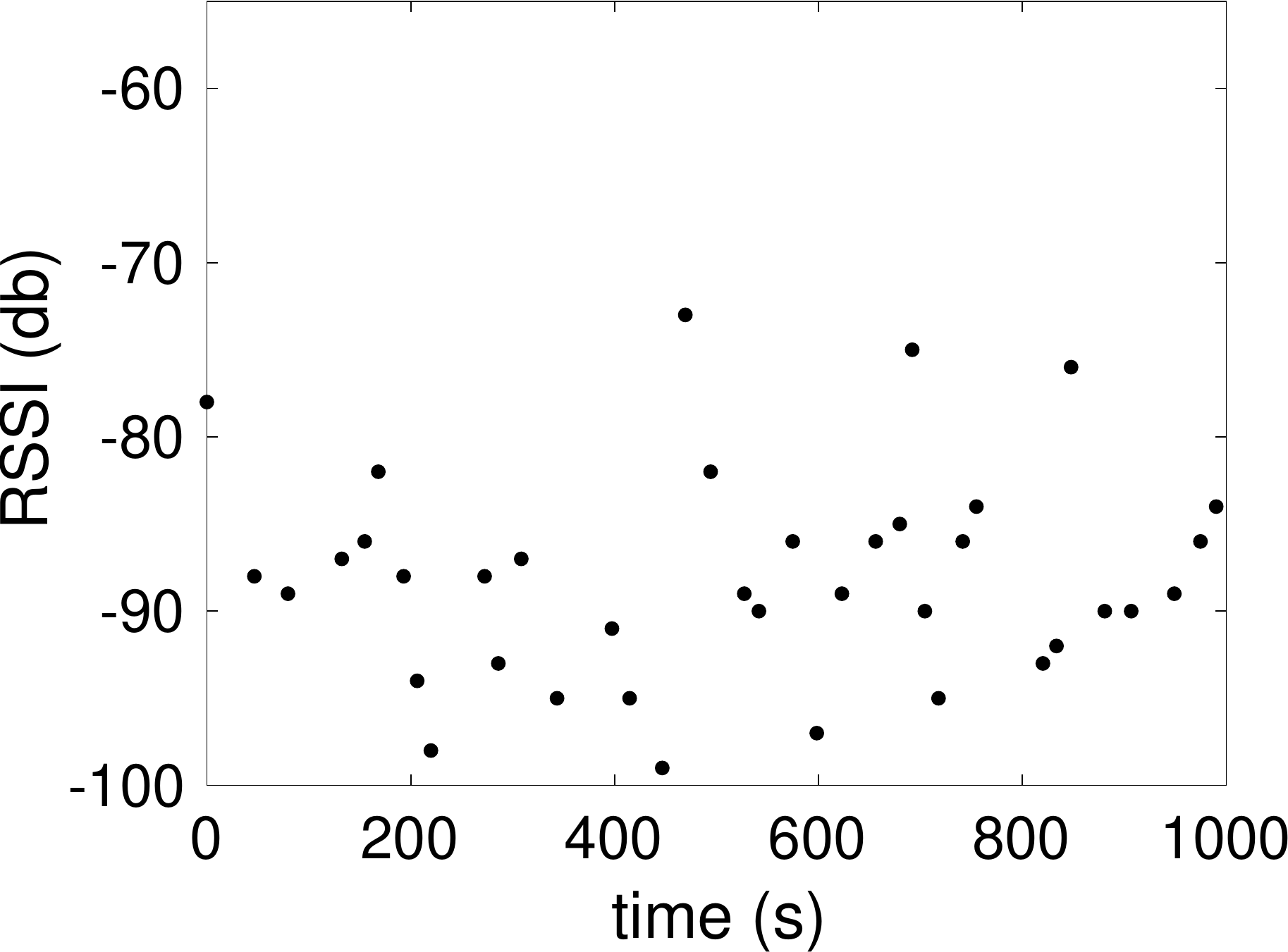}
}
\subfloat[2m behind]{
\includegraphics[width=0.45\columnwidth]{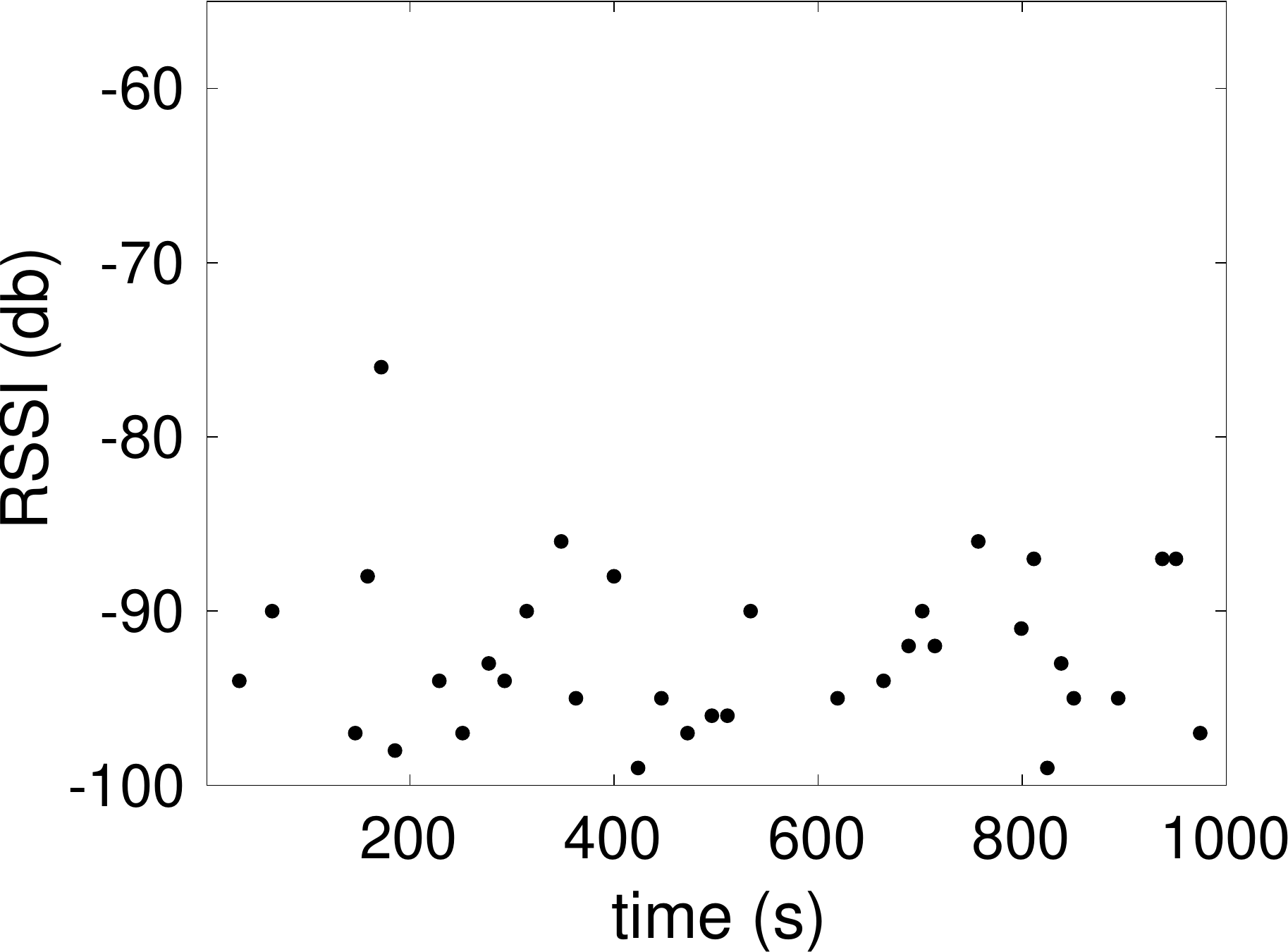}
}
\caption{Measurements of received signal strength for two people carrying mobile handsets and walking in four different configurations. In (a) the two people are walking side by side (shoulders touching), in (b) side by side but with 1m distance maintained between them.  In (c) the people are walking one in front of the other with a 1m gap and in (d) with a 2m gap. }\label{fig:walking}
\end{figure}

Figure \ref{fig:walking} shows time histories of the measured received signal strength for each configuration.  When the two people are walking close together, Figure \ref{fig:walking}(a), the received signal strength is around -65dB$\pm$10dB.   From Figure \ref{fig:walking}(d) it can be seen that when walking with a 2m gap the received signal strength consistently falls to around -95dB$\pm$10dB.   That is, there is a clear shift in received signal strength as the distance changes.  This suggests that the limited task of distinguishing between whether people outdoors are side by side or one is 2m behind the other, when no other configurations can occur, can indeed likely be achieved using Bluetooth LE received signal strength data.   

Unfortunately the situation becomes more complex when other configurations are considered.   Figure \ref{fig:walking}(b) shows measurements taken when walking side by side while maintaining a 1m gap and Figure \ref{fig:walking}(c) when walking one behind the other with a 1m gap.  In the first case the received signal strength is around -75db$\pm$10dB but in the second case it is much lower at around -92dB$\pm$10dB i.e. similar to the signal strength measured when one behind the other and 2m apart.   This indicates that this data cannot readily be used to distinguish between whether people are 1m or 2m apart when they are walking behind each other.   That is, this data suggests that we cannot reliably distinguish whether people are located less than 2m of each other when walking behind each other in a city street, although we may be able to distinguish this when people are walking side by side. 

\begin{figure}
\centering
\includegraphics[width=0.9\columnwidth]{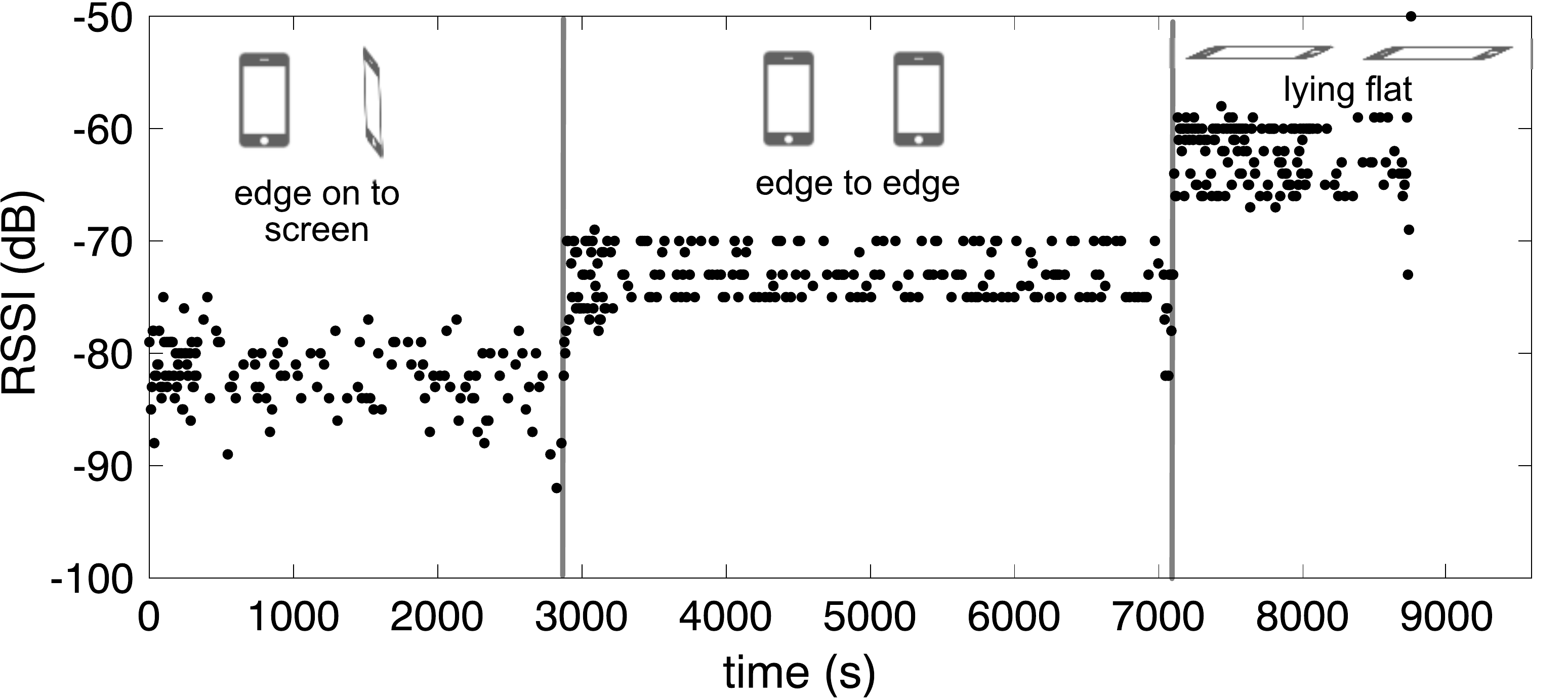}
\caption{Impact of handset orientation on received signal strength.   Handsets are placed 1m apart and the received signal strength recorded when (a) one handset is oriented edge on to the screen of the other handset, (b) when both handsets are oriented edge on to one another and (c) when both handsets are lying flat.  These changes in orientation result in a change in received signal strength of around 20dB.}\label{fig:orientations}
\end{figure}

Further investigation suggests that the substantial difference in measured received signal strength at a distance of 1m seen in Figures \ref{fig:walking}(b) and \ref{fig:walking}(c) is likely due to the changes in the relative orientations of the handsets when walking side by side compared to when walking one behind the other.   To help gain more insight into this effect Figure \ref{fig:orientations} shows measurements taken with two handsets placed 1m apart in fixed positions within an open indoor area.   The relative orientations of the handsets are adjusted and the measured received signal strength recorded.  It can be seen that when one handset is edge on to the screen of the other (so the two handsets are at a 90$^\circ$ angle to one another) the received signal strength fluctuates around -85dB.    When the handsets are adjusted to be aligned edge on to one another the received signal strength increases by about 10dB to around -75dB, and when the handsets are then both placed face down the received signal strength increases again by about 10db to around -65dB.   

Roughly speaking, when two people are walking one behind the other with handsets in their pockets then the handsets are orientated edge on, i.e. similarly to the configuration in the middle section of Figure \ref{fig:orientations}.   When the people walk side by side the handsets are roughly orientated so that the screens face one another.  Data for this configuration is not shown in Figure \ref{fig:orientations}, but our measurements indicate that the received signal strength at 1m is similar to that when the phones are lying flat i.e. to the right-hand side of Figure \ref{fig:orientations} and around 10dB higher than when the handsets are edge on.   The data in Figure \ref{fig:orientations} is for a controlled indoor setup with no obstructions from people's bodies etc and so is not directly comparable to the data in Figures \ref{fig:walking}(b)-(c), but it does suggest that significant shift in received signal strength observed when walking side by side vs one behind the other can largely be explained by the change in relative phone orientation.

\subsection{Scenario 2: Sitting Around A Meeting Table}
Our second scenario aims to evaluate proximity measurement within an office-based workplace.  A crude model of workplace movement is that during the work day people mainly spend time either (i) at their desk and (ii) in meetings.  Regarding (i), if it is a shared office then if one person becomes infected their office mates are known and so contact-tracing is straightforward (with the possible exception of a large open-plan office, but we leave evaluation of that more complex scenario to future work).  Regarding (ii), the hope is that to assist with contact tracing we can augment an infected persons recollection of meetings attended and of the other people present by using Bluetooth LE received signal strength measurements.   We assume that during a meeting people spend the bulk of their time sitting around a table and so we try to evaluate the accuracy of Bluetooth LE signal strength data for proximity detection in this scenario..

\begin{figure}
\centering
\includegraphics[width=0.3\columnwidth]{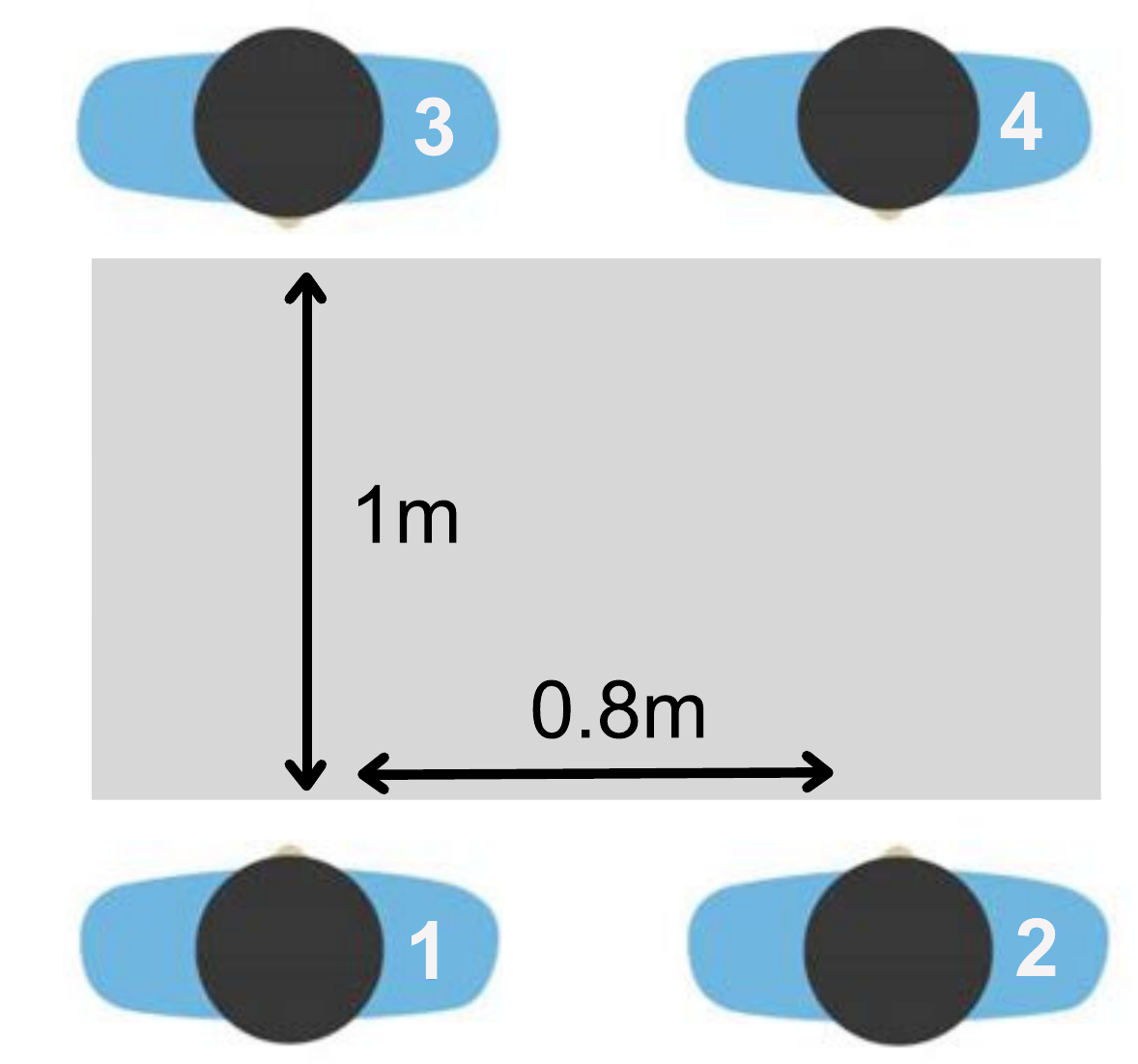}
\caption{Experimental setup: arrangement of four people around table.}\label{fig:table_setup}
\end{figure}

We collect measurements of Bluetooth LE received signal strength with four people sitting around a wooden table as illustrated schematically in Figure \ref{fig:table_setup}.  We take measurements both when people have their mobile handset in their trouser pocket and when it is placed on the table.

\begin{figure}
\centering
\subfloat[Person 1 $\leftrightarrow$ Person 2]{
\includegraphics[width=0.45\columnwidth]{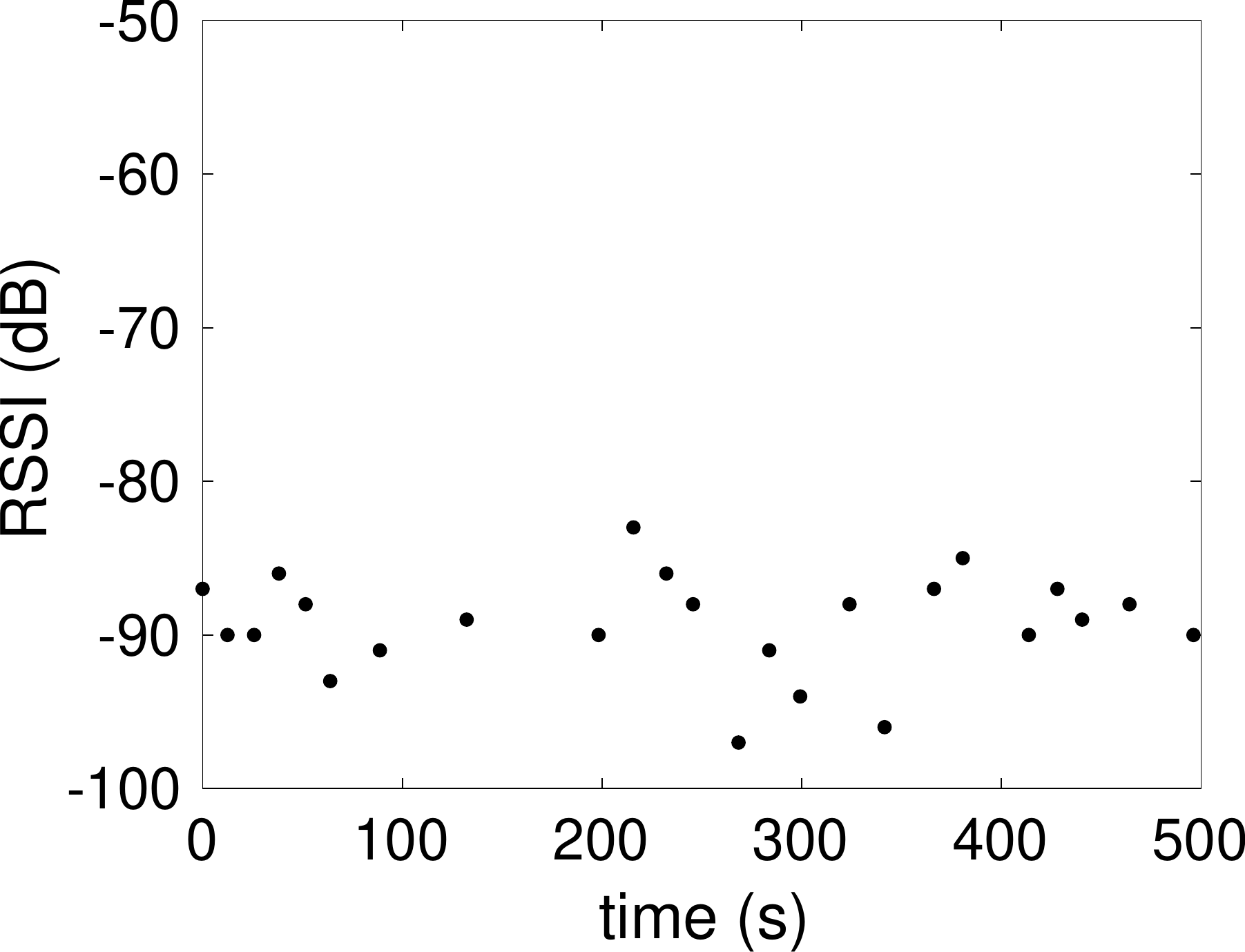}
}
\subfloat[Person 1 $\leftrightarrow$ Person 3]{
\includegraphics[width=0.45\columnwidth]{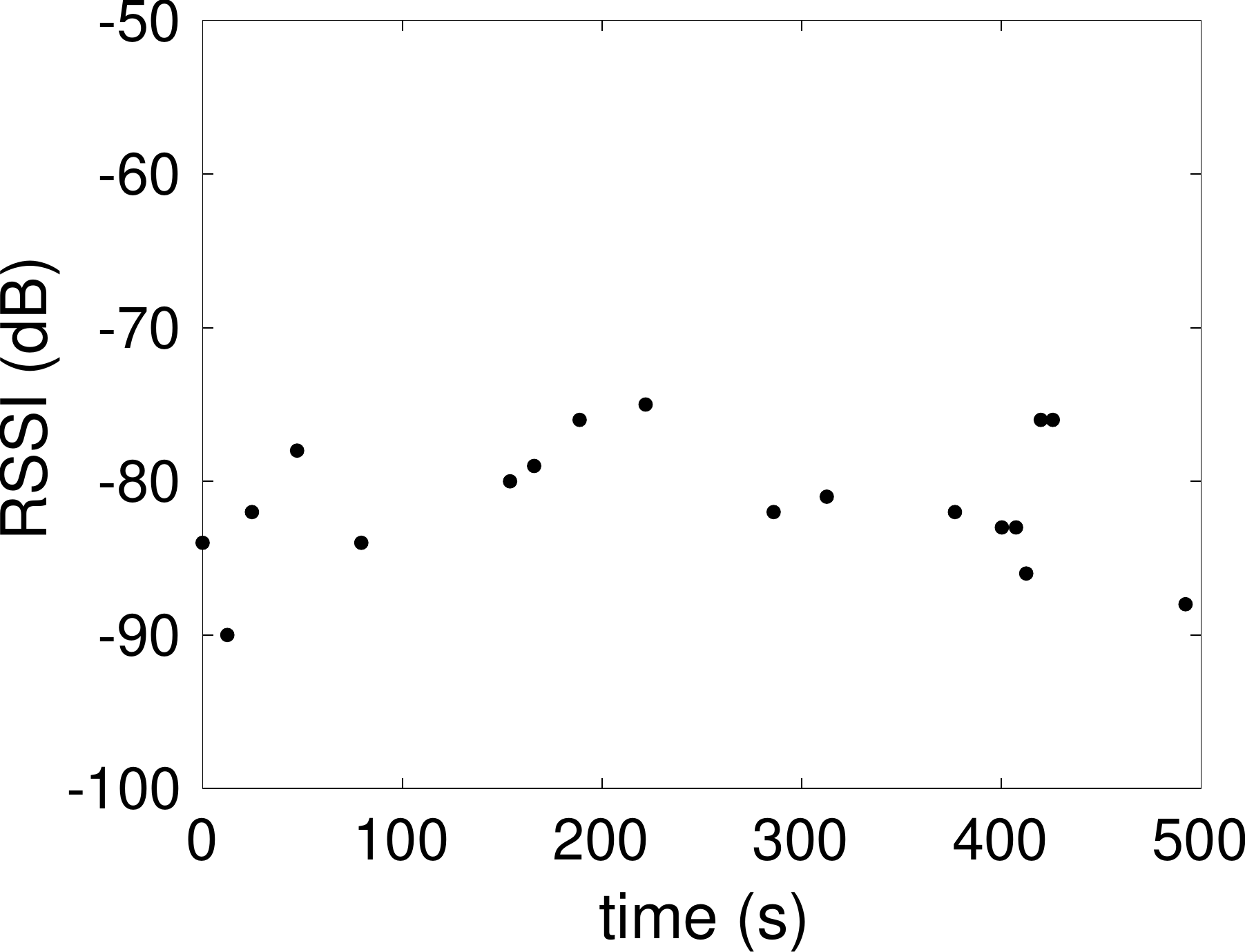}
}
\vspace{0.2cm}
\subfloat[Person 1 $\leftrightarrow$ Person 2]{
\includegraphics[width=0.45\columnwidth]{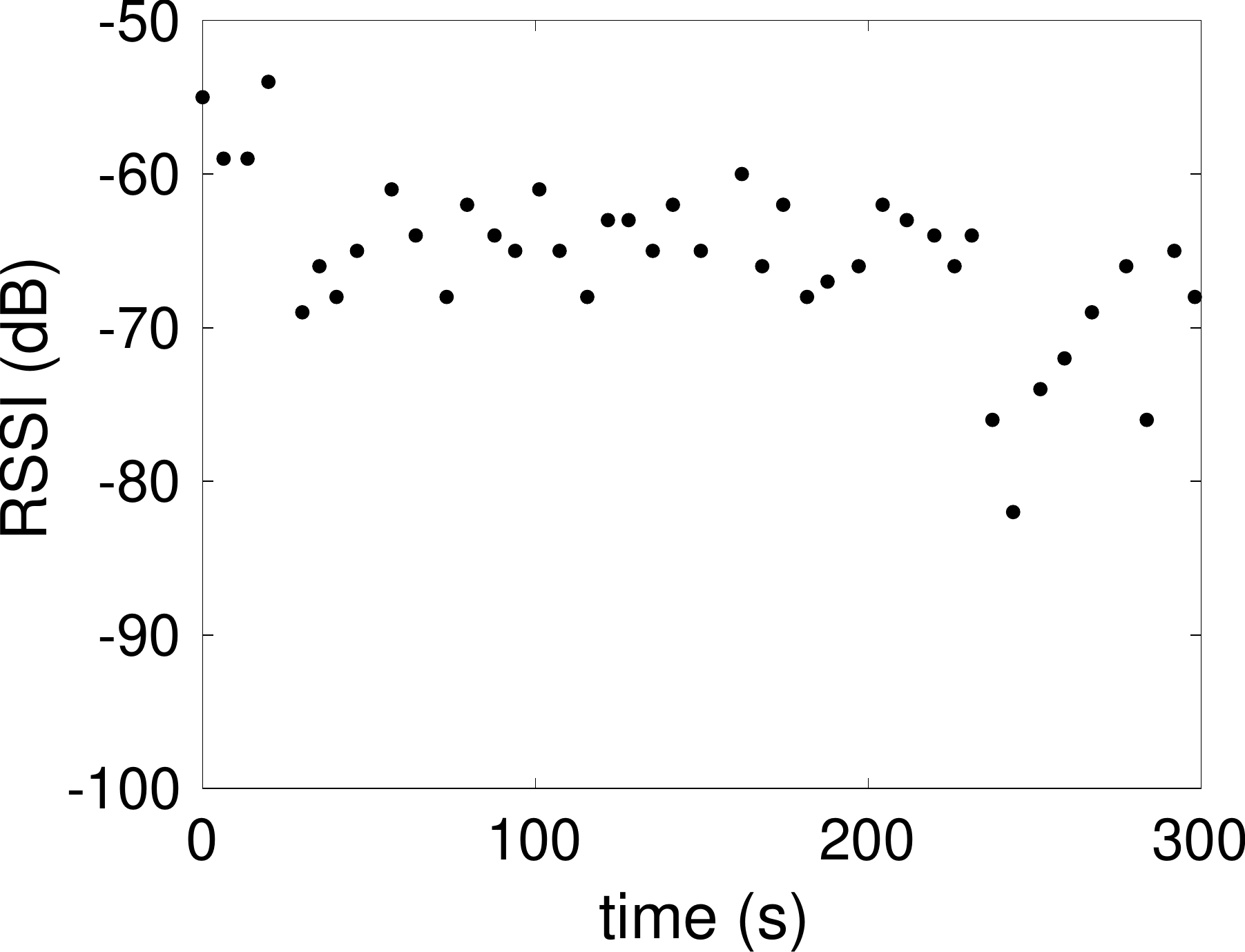}
}
\subfloat[Person 1 $\leftrightarrow$ Person 3]{
\includegraphics[width=0.45\columnwidth]{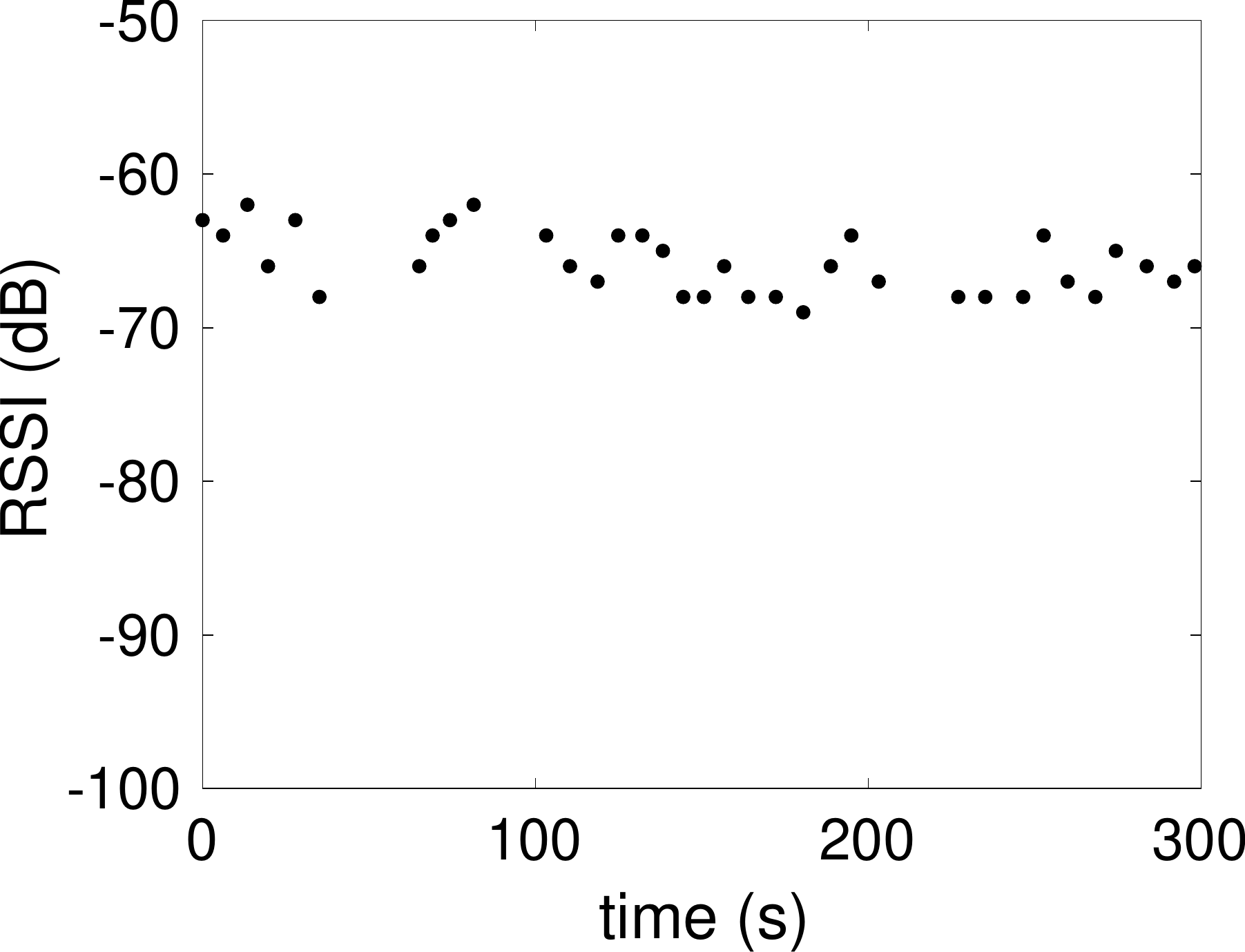}
}
\caption{Measurements of received signal strength for four people sitting around a wooden table, see Figure \ref{fig:table_setup}.   In (a) and (b) the mobile handset is placed in the persons trouser pocket, in (c) and (d) the handsets are placed on the table in front of the person.}\label{fig:table}
\end{figure}

Figures \ref{fig:table}(a)-(b) shows the received signal strength measured between person 1 and person 2 (i.e. two people sitting beside each other at the table) and between person 1 and person 3 (i.e. between people sitting opposite each other) when their mobile handset is in their trouser pocket.   Figures \ref{fig:table}(c)-(d) show the corresponding data when each person places their mobile handset on the table in front of them.   It can seen that there is a substantial difference in signal strength between situations where the handsets are in people's pockets vs when they are placed on the table.   When placed on the table the received signal strength is around -65dB, a relatively high level that allows the inference that the two people are located close together.   However, when the handsets are in people's trouser pockets the received signal strength is much lower at around -80dB to -90dB.   The low signal strength observed when the handsets are in people's trouser pockets is caused by a combination of signal absorption by people's bodies and the relative orientations of the phones.

Based on this received signal strength data this suggests it would be hard to detect when people sitting close together at a table unless people place their handsets on the table during the meeting.
 

\begin{figure}
\centering
\subfloat[Sitting side by side]{
\includegraphics[width=0.45\columnwidth]{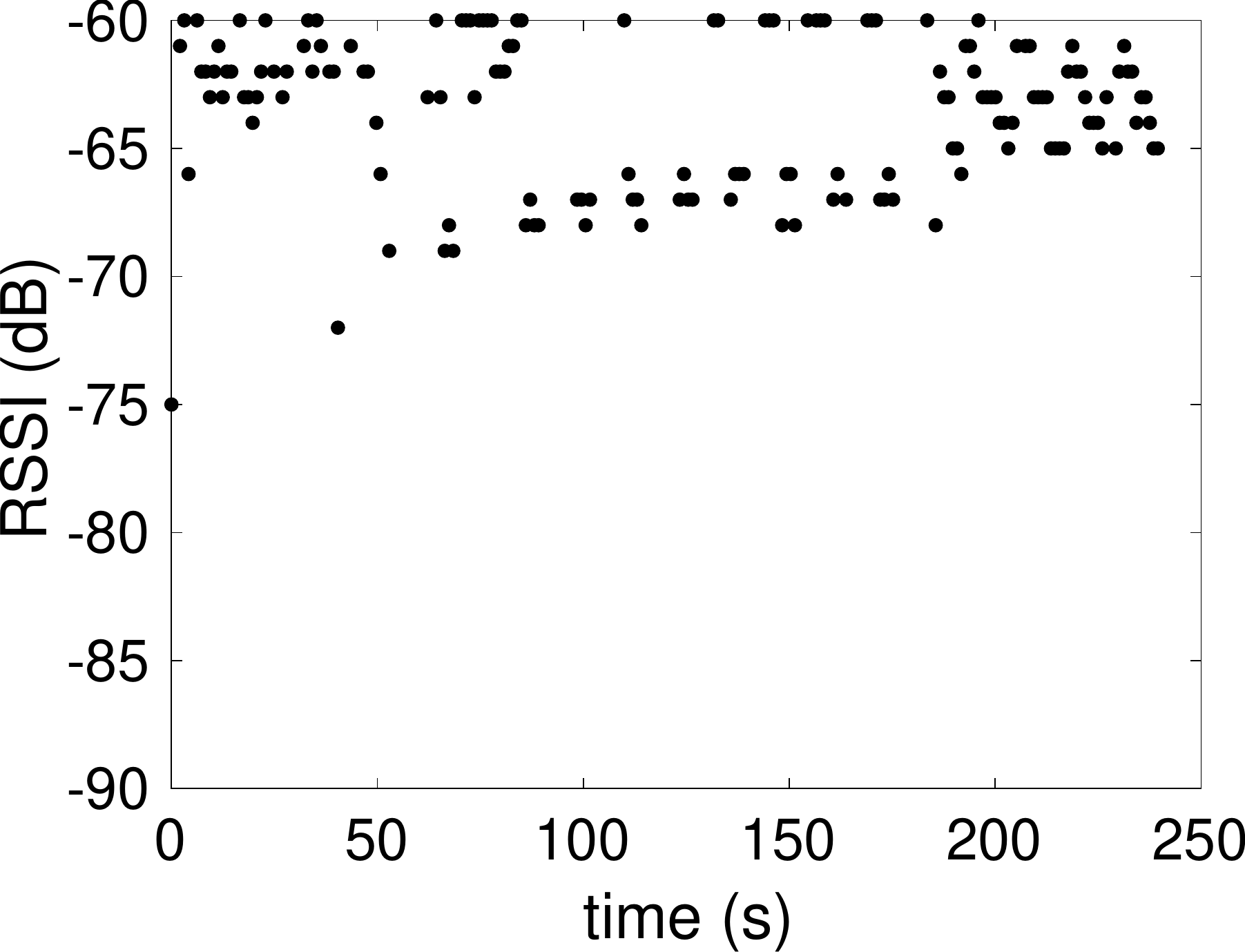}
}
\subfloat[Train layout]{
\includegraphics[width=0.45\columnwidth]{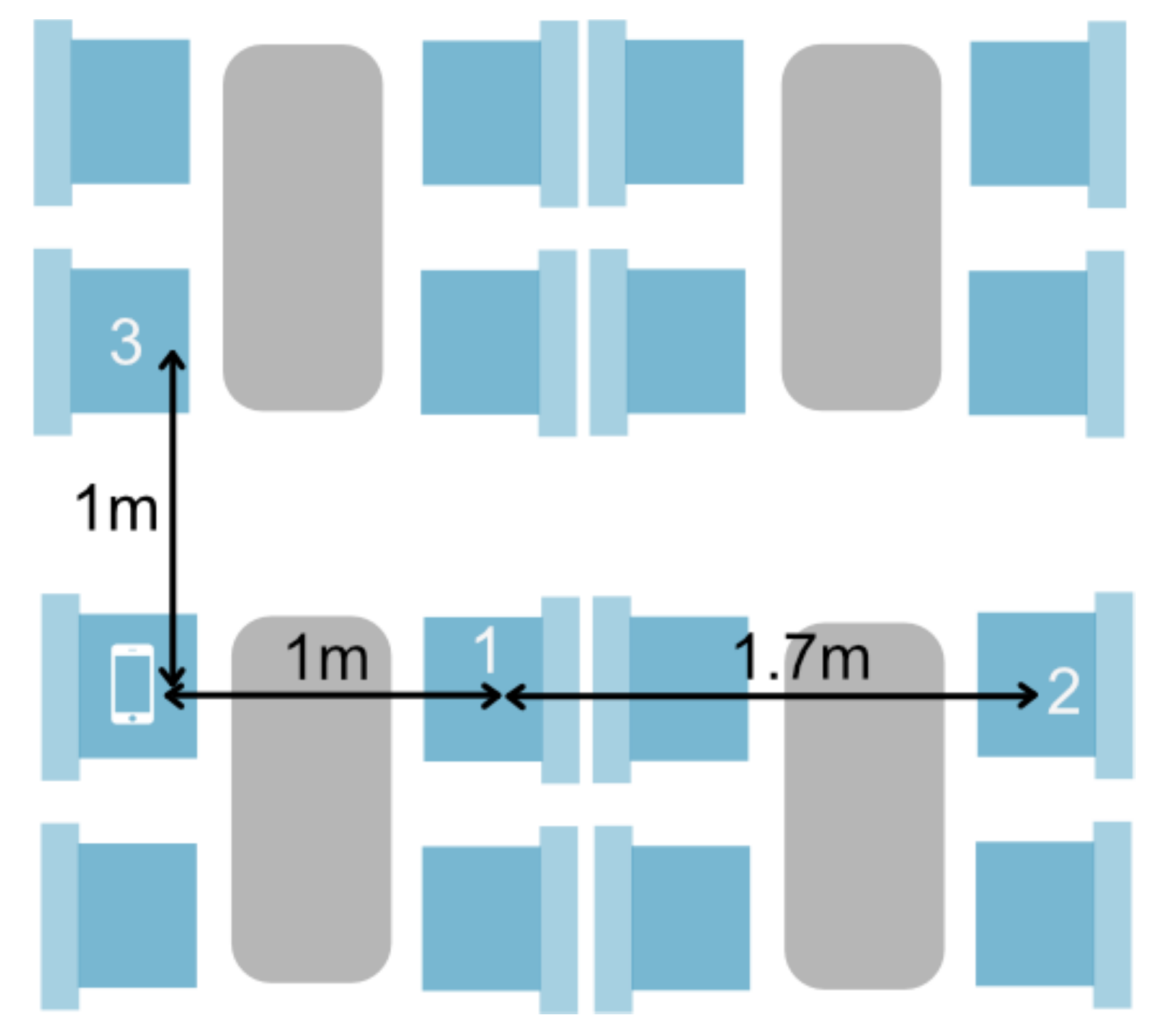}
}
\vspace{0.2cm}
\subfloat[Sitting two rows behind]{
\includegraphics[width=0.45\columnwidth]{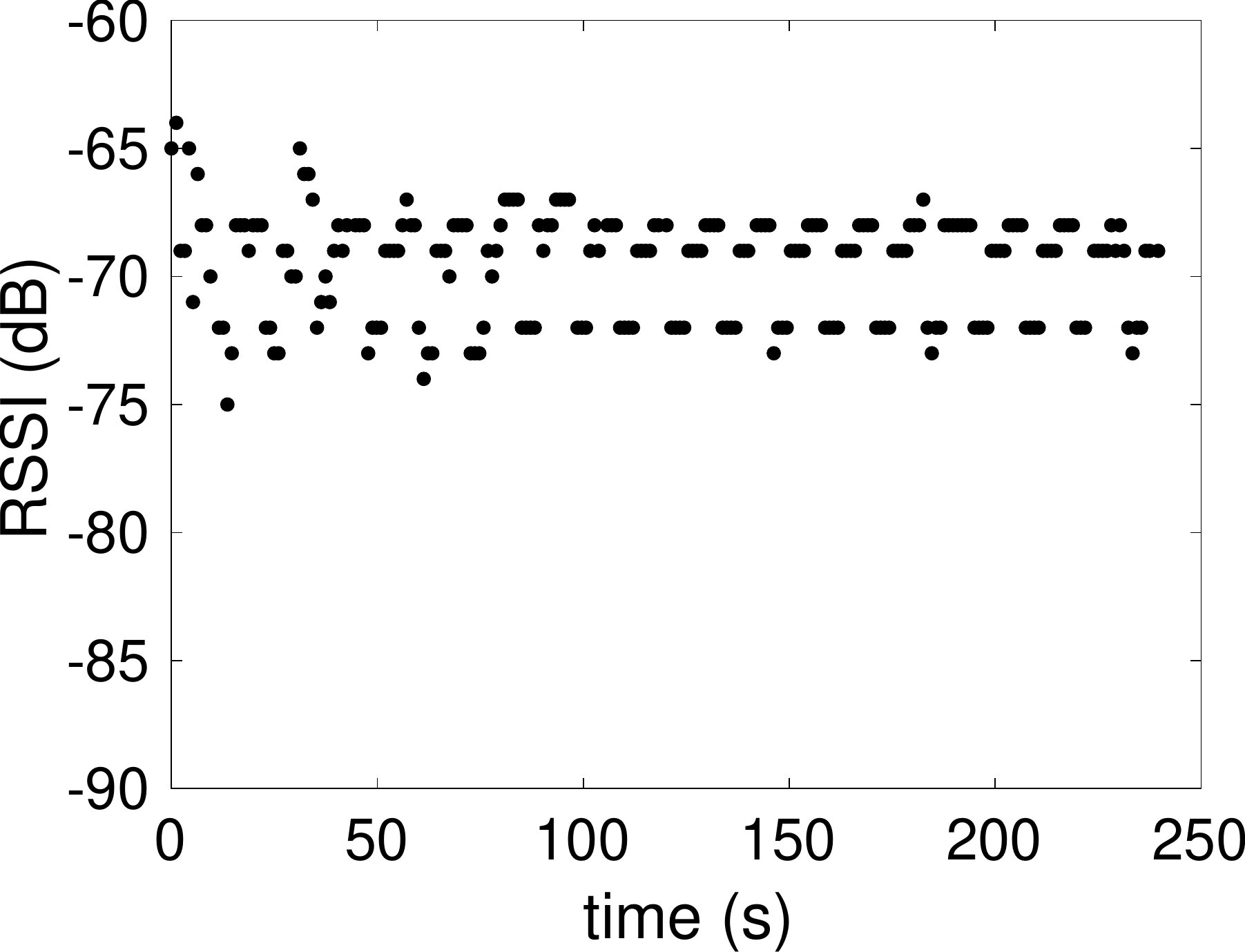}
}
\subfloat[Sitting row opposite]{
\includegraphics[width=0.45\columnwidth]{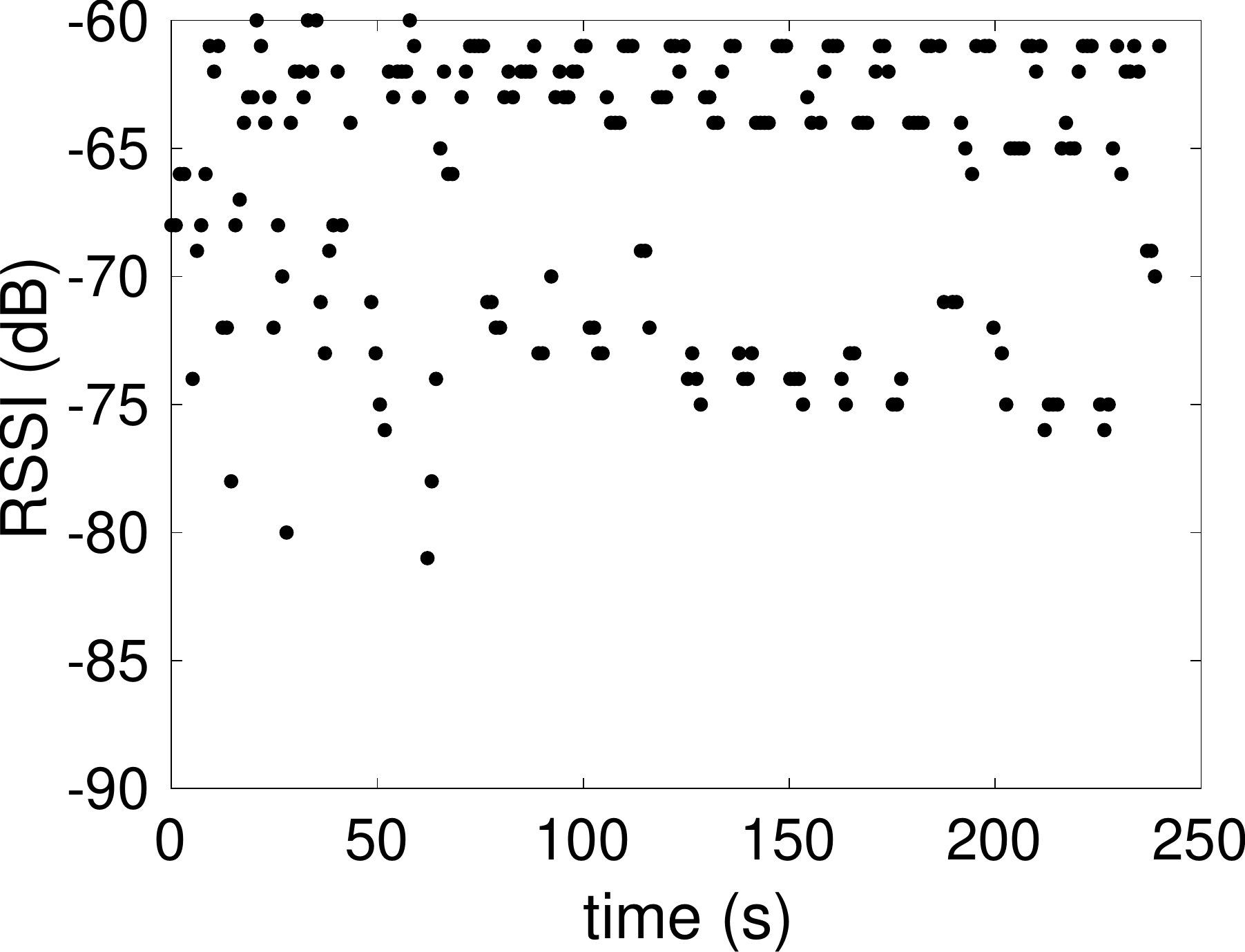}
}
\caption{Measurements of received signal strength inside a train carriage.  A handset is placed in one seat as indicated in schematic (b) and the received signal strength measured between this and a handset located in a person's trouser pocket when they are sitting in the positions marked 1, 2 and 3 in schematic (b).  Plot (a) shows measurements from position 1, (c) from position 2 and (d) from position 3. }\label{fig:train_sligo}
\end{figure}

\subsection{Scenario 3: Sitting In A Train Carriage}
Our third scenario aims to evaluate proximity measurement while travelling on public transport.    Our rough model is that people mainly spend their time seated and so we take measurements of the received signal strength between various seating positions.   We focus on a train since Irish Rail very kindly let us take measuments on two of their rail carriages at short notice, but of course buses, trams and aircraft are also important.   While our measurements are taken within a stationary train carriage we do not expect movement of the carriage to change things much.

Figure \ref{fig:train_sligo}(b) shows the carriage layout.  Seats are arranged in groups of four around a small table.  We placed a handset on one seat, indicated at the bottom left of Figure \ref{fig:train_sligo}(b).  A person with a second handset in their left trouser pocket then sat in the positions marked 1, 2 and 3 in Figure \ref{fig:train_sligo}(b) and the received signal signal strength from the first handset recorded.   Figures \ref{fig:train_sligo}(a), (d) and (d) show the measurements obtained.   It can be seen from Figures \ref{fig:train_sligo}(a) and \ref{fig:train_sligo}(c) that the received signal strength is about -62dB when the person is seated around 1m from the first handset, which is in quite good agreement with Figure \ref{fig:garden}(a).  When seated around 2.7m from the first handset the signal strength falls to about -70dB, also in reasonable agreement with Figure \ref{fig:garden}(a).   

\begin{figure}
\centering
\subfloat[RSSI vs distance]{
\includegraphics[width=0.45\columnwidth,valign=t]{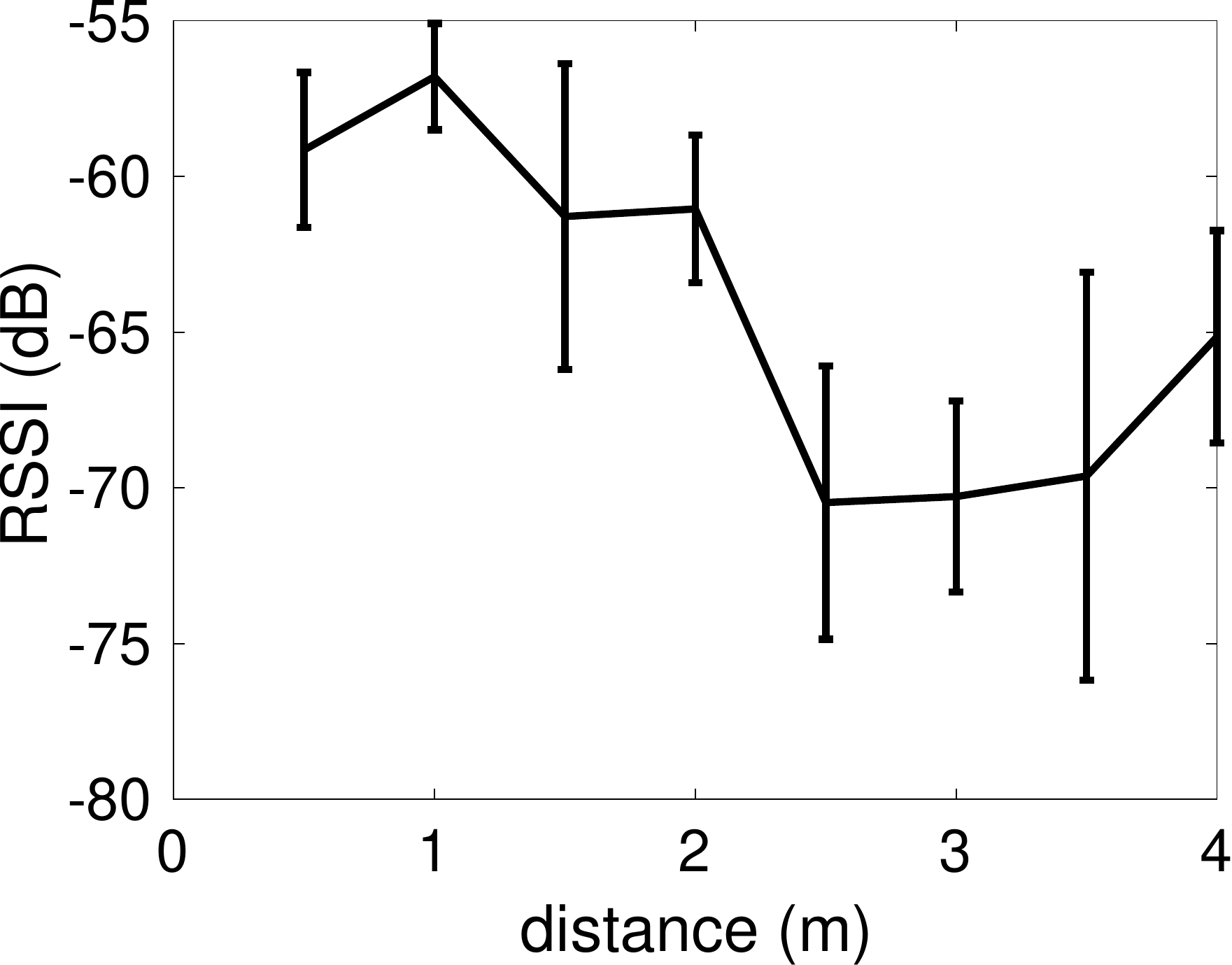}
}
\subfloat[Train carriage setup]{
\includegraphics[width=0.39\columnwidth,valign=t]{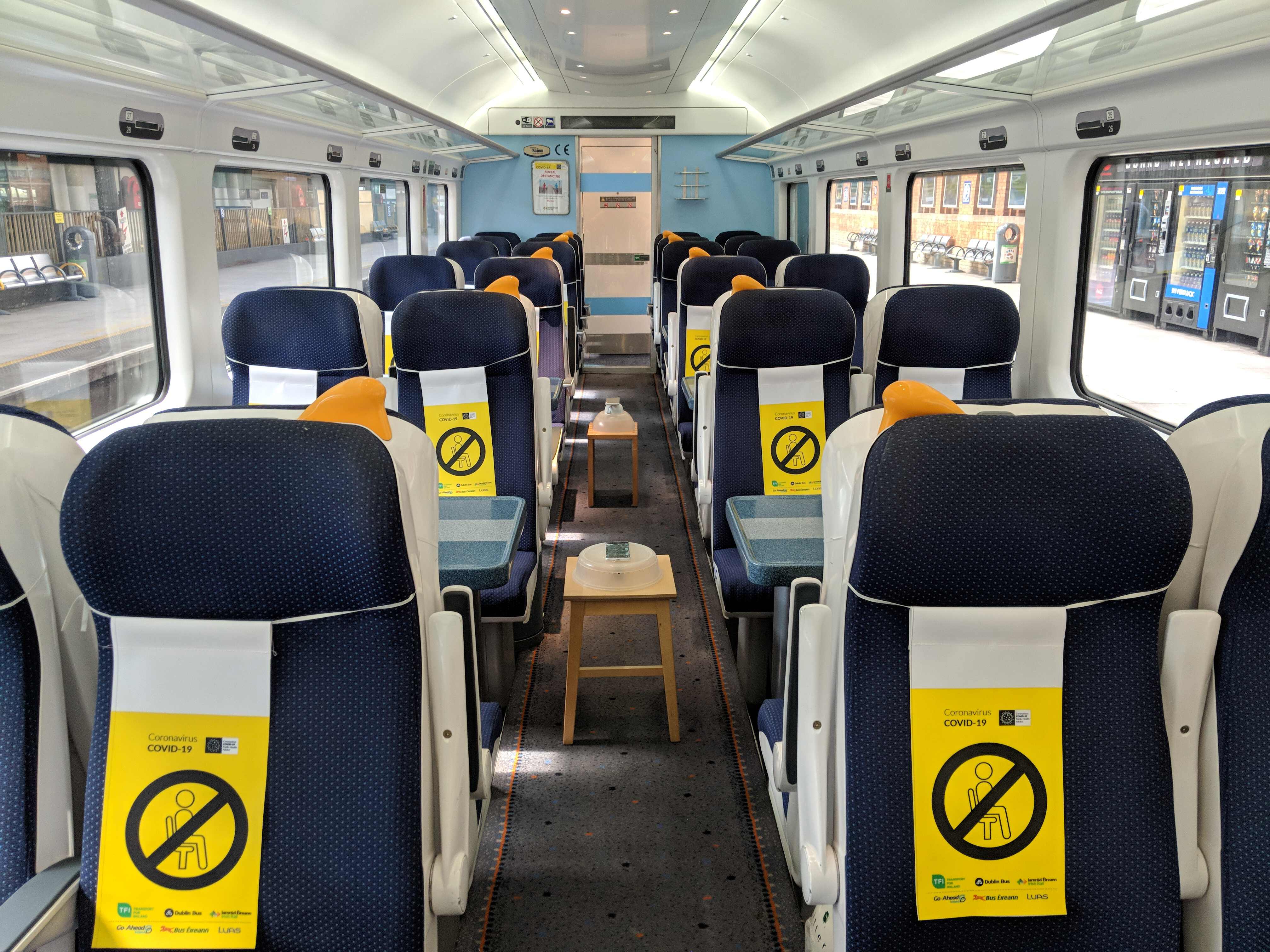}
}
\caption{(a) Measured received signal strength vs distance along the aisle inside the train carriage shown in (b).   }\label{fig:train_dist}
\end{figure}

We also collected measurements of received signal strength vs distance along the aisle in the middle of the rail carriage, roughly at the seat height (approximately 50cm above the floor), see Figure \ref{fig:train_dist}(b).  The measurement setup used is the same as that used in Figure \ref{fig:garden}.   Figure \ref{fig:train_dist}(a) shows the measured received signal strength, with the error bars indicating one standard deviation.  It can be seen that the received signal strength remains roughly constant up to a distance of 2m and then falls sharply.  As noted above, seats in this carriage are arranged in groups of four around small tables.  The sharp fall in received signal strength coincides with moving from one group of four seats to another.   This data therefore suggests that the signal strength is high between seats with the same group but lower between seats in different groups.   

It can be seen from Figure \ref{fig:train_dist}(a) that the received signal strength then stays roughly constant out to a distance of 4m, but appears to \emph{increase} when moving from 3.5 to 4m (recall that we expect signal strength to generally fall with increasing distance).   Hence while our measurements suggest that received signal strength might be used to distinguish between people sitting in the same group of seats and those sitting in a different group, the increase in signal strength at 4m is potentially of concern for proximity detection based on received signal strength.    We note that the walls, floor and ceiling of a train carriage are primarily made of metal, albeit with the walls and ceiling lined with plastic cladding and the floor with carpet.   The seats also likely contain metal.  Since metal strongly reflects radio signals it is unsurprising that the radio signal propagation behaves in quite a complex manner and a further, more extensive, measurement campaign would be prudent.     


%


\subsection{Scenario 4: Grocery Shopping}

This scenario aims to evaluate the use of Bluetooth LE for proximity measurement while grocery shopping in a typical Irish supermarket.   Grocery shopping is, of course, something that everyone has to do and under the current lockdown for many people it is also one of the few activities where they are likely to come into relatively close contact with people outside their immediate social circle.

\begin{figure}
\centering
\subfloat[Supermarket location]{
\includegraphics[width=0.39\columnwidth]{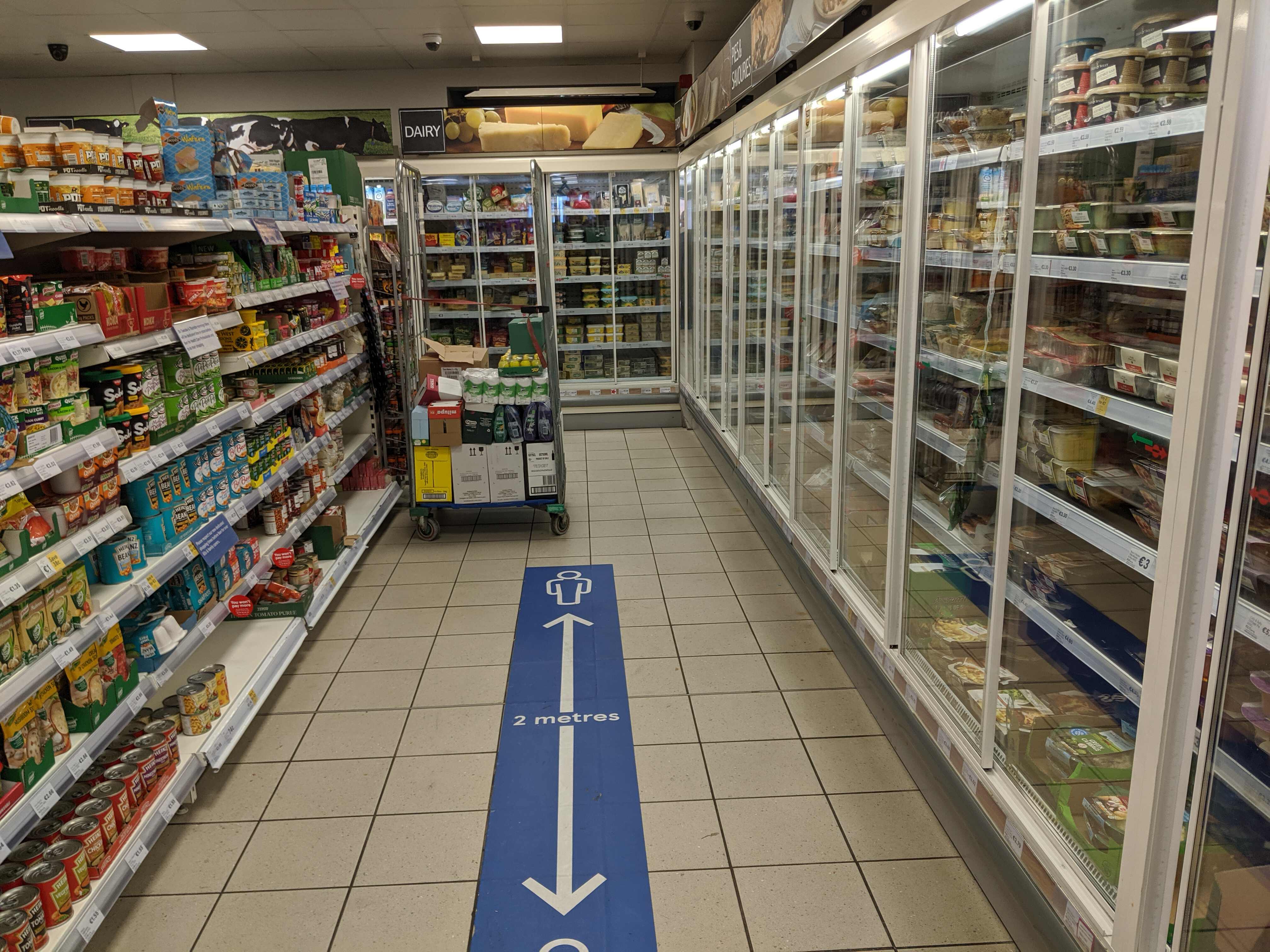}
}\\
\subfloat[Close together]{
\includegraphics[width=0.45\columnwidth]{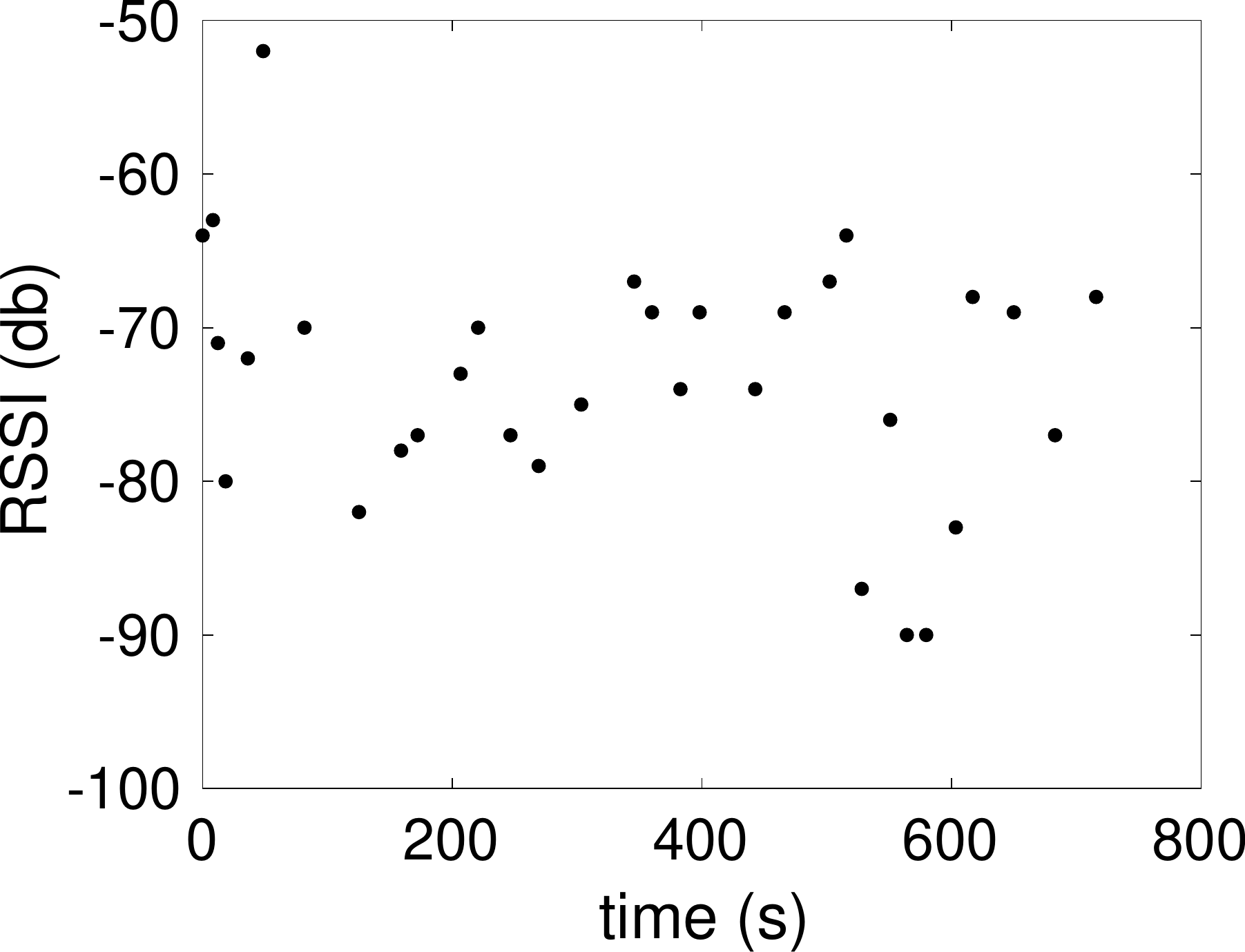}
}
\subfloat[2m apart]{
\includegraphics[width=0.45\columnwidth]{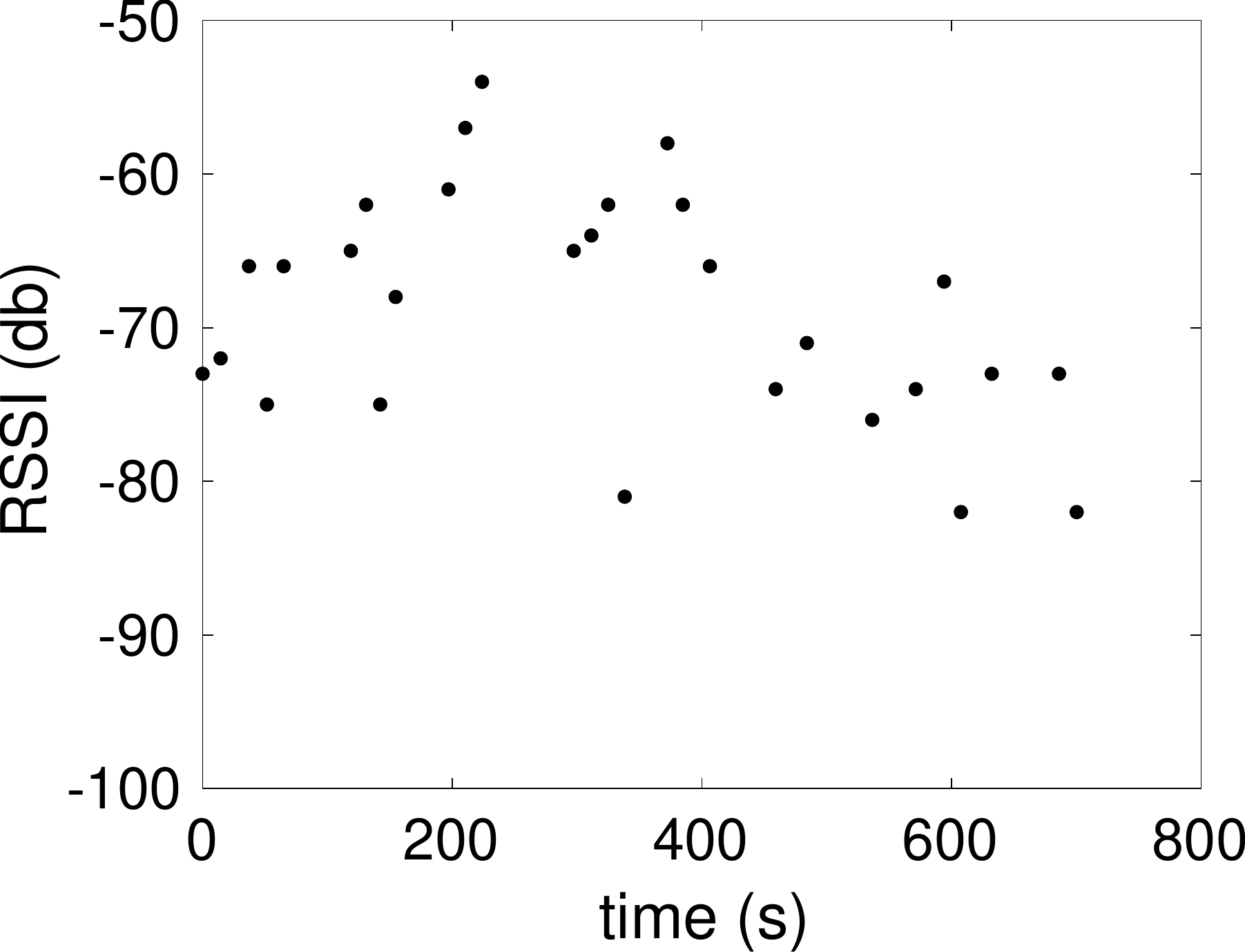}
}
\caption{Measurements of  received signal strength for two people walking around a supermarket with a shopping trolley.  In (b) they stay close together while walking around the aisles for 10mins, in (c) they walk one behind the other maintaining a distance of 2m.}\label{fig:tesco}
\end{figure}

Figure \ref{fig:tesco} shows received signal strength measurements taken while two people were walking around a large supermarket in suburban Dublin pushing a standard metal shopping trolley.   The supermarket is organised into aisles that are around 2m wide, with metal shelves, fridges/freezers etc and other shoppers (who are social distancing) are also present, see Figure \ref{fig:tesco}(a).  The two people walk one behind the other (the relatively narrow aisles in the shop encourage this in any case) and both carry a phone in their left-hand trouser pocket.  

In Figure \ref{fig:tesco}(b) the two people stay close together (although still one behind the other) while walking, in Figure \ref{fig:tesco}(c) they walk one behind the other maintaining a distance of 2m.   Somewhat surprisingly, it can be seen that the received signal strength measurements look much the same in both cases.   

Presumably this is due to a mix of the impact of the handset orientation already noted when walking one behind the other, and of the effect on wireless signal propagation of the complex environment (metal shelves etc) within the supermarket.   

These measurements suggest that received signal strength probably cannot be used to distinguish between whether people are close together or 2m apart when walking around a supermarket.   However, we note that it is probably unlikely that people will spend more than 15 minutes within 2 metres of each other while walking around a supermarket, and so failure to detect close proximity using received signal strength is perhaps of less importance than in other scenarios.   Of more concern are false positives, where people are detected as being close together when in fact they are not, in light of the relatively high received signal strengths we see in Figure \ref{fig:tesco}(c) when people are 2m apart.

\begin{figure}
\centering
\subfloat[2m apart]{
\includegraphics[width=0.45\columnwidth]{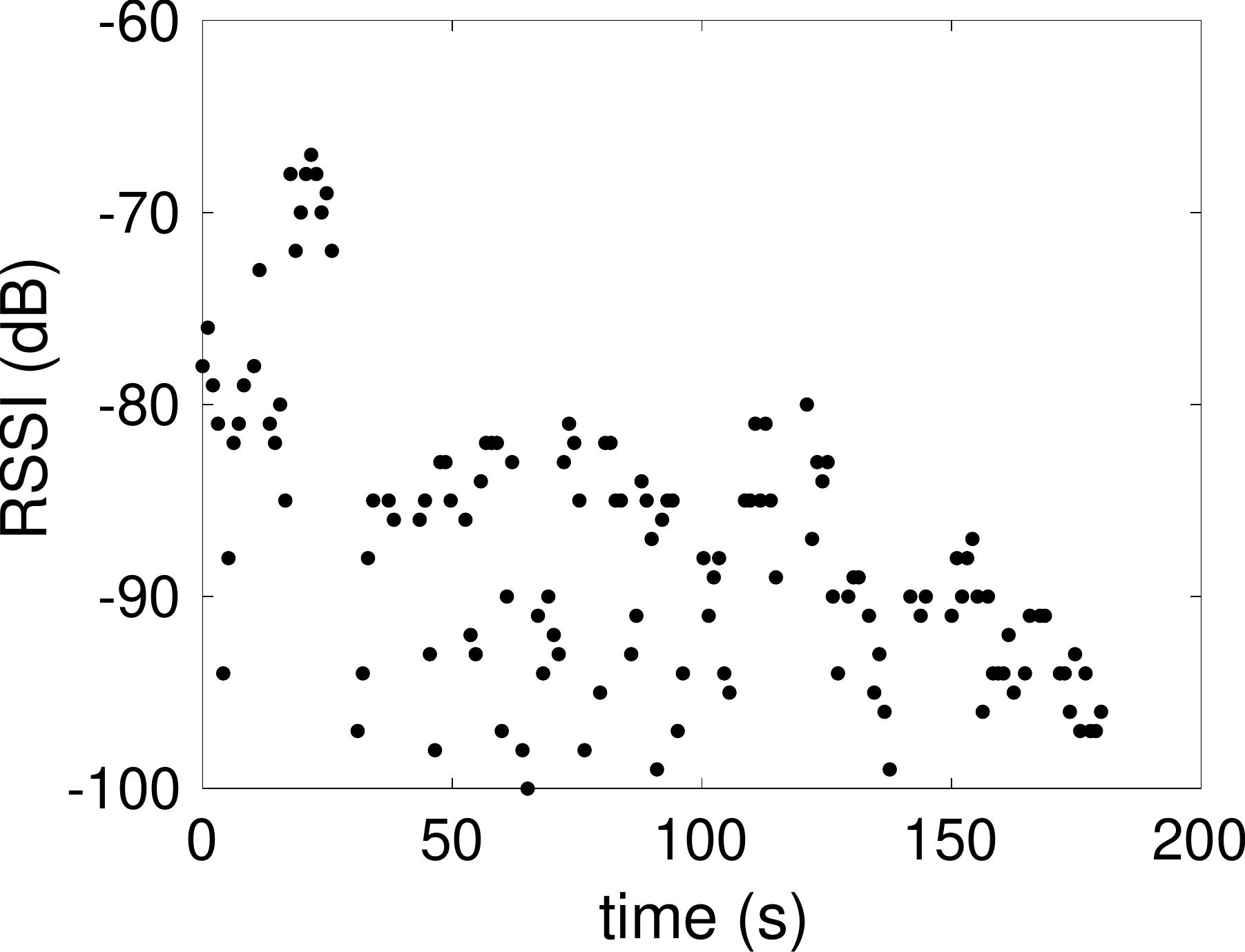}
}
\subfloat[4m apart]{
\includegraphics[width=0.45\columnwidth]{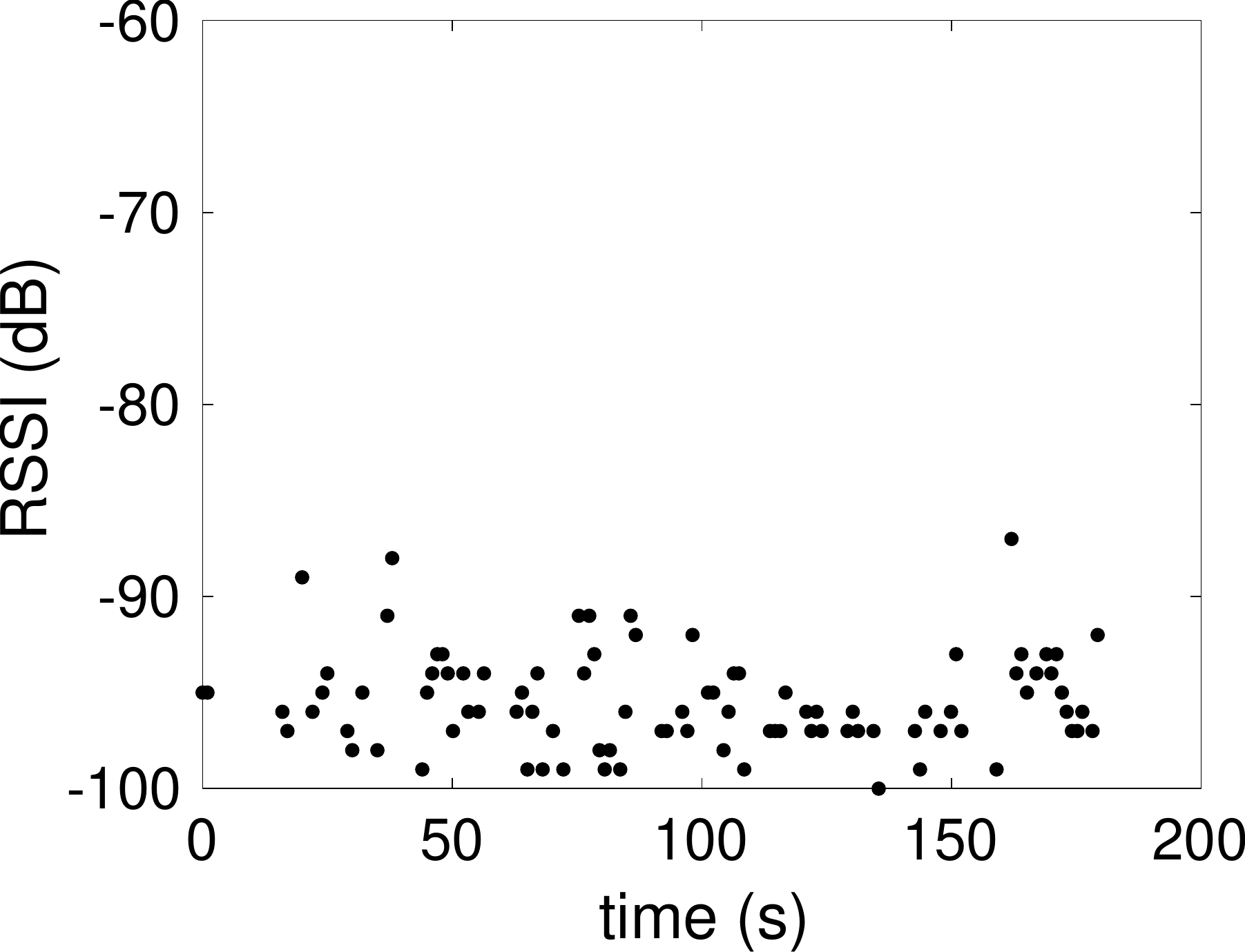}
}\\
\subfloat[4m apart, third person between]{
\includegraphics[width=0.45\columnwidth]{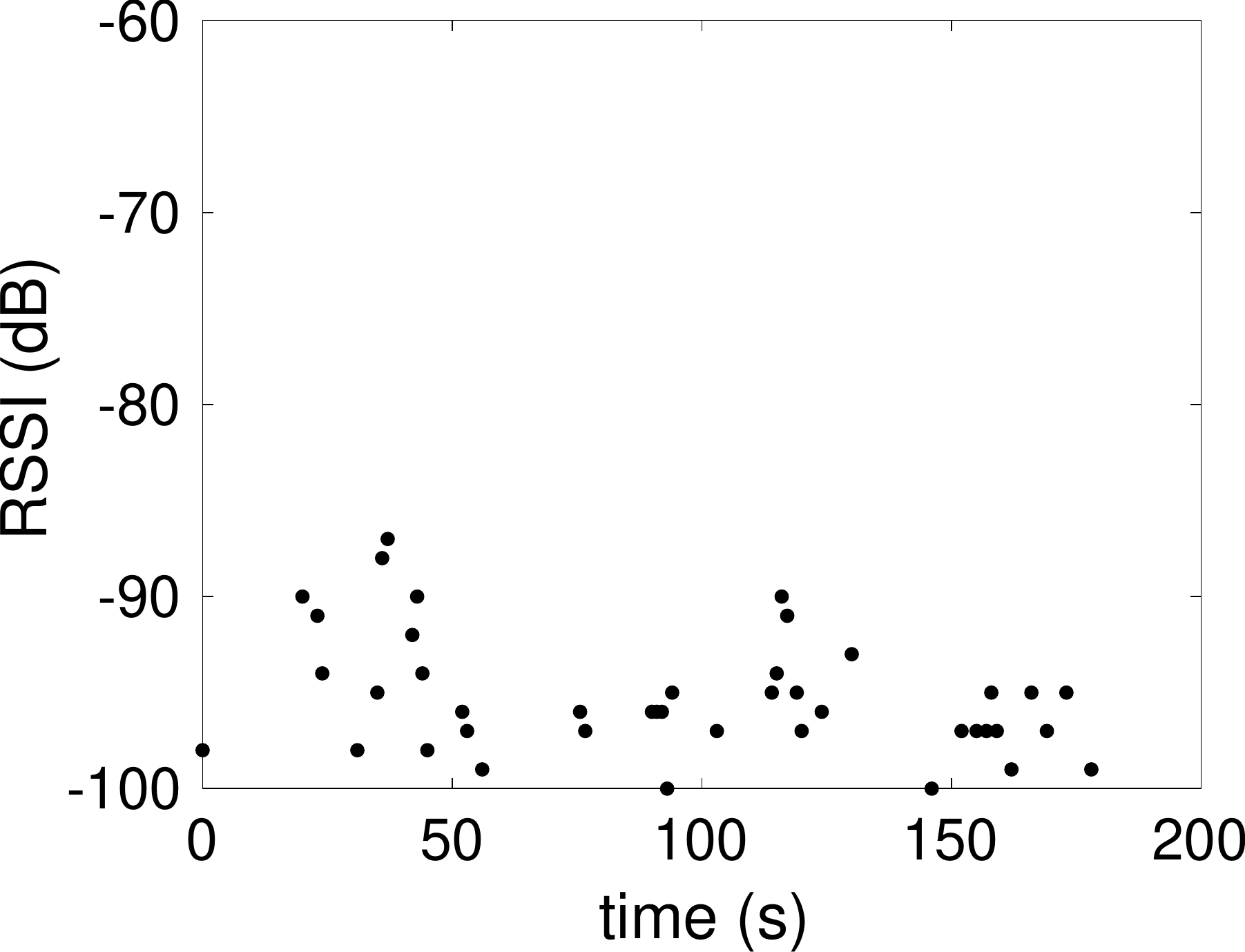}
}
\caption{Measurements of  received signal strength for people queueing outside a supermarket.  In (a) they are 2m apart, in (b) 4m apart and in (c)  they are 4m apart but with a third person standing midway i.e. a queue of 3 people each standing 2m apart.  Measurements taken using a pair of Samsung Galaxy A10s.}\label{fig:lidl}
\end{figure}

Current social distance measures in Ireland limit the number of people who can be inside a shop at the same time.  This means that people often need to queue outside until allowed to enter.  While queueing people are asked to remain spaced 2m apart.  Figure \ref{fig:lidl} shows received signal strength measurements collected while queueing outside a supermarket beside metal shopping trolleys.   Comparing Figures \ref{fig:lidl}(a) and \ref{fig:lidl}(b) it can be seen that the received signal strength falls as people stand further apart.   A third person standing in between does not change the received signal strength by much, see Figure \ref{fig:lidl}(c).    This data suggests that received signal strength might successfully be used to detect proximity while queueing.

%

\section{Related Work}
While there have been some studies on use of the magnetometers on smartphones
for contact tracing, e.g. see~\cite{magnetometers2017,magnetometers2019}, to
the best of our knowledge there are no previous measurement studies on the use
of Bluetooth LE specifically in the context of contact tracing. 
Perhaps the closest work is FluPhone~\cite{yoneki2011fluphone},
an approach proposed in 2011 that made use of Bluetooth rather than Bluetooth LE and did not try to distinguish between contact events where people are less than 2m apart and when they are further away (the app simply logged all observed Bluetooth MAC addresses).  Use of Bluetooth LE for proximity detection has, however, been investigated in the context of measuring social interaction and also indirectly in the context of
indoor localisation.
 
\subsection{Proximity Detection Using Bluetooth LE} Bluetooth LE was
standardised in 2010 with the first devices using Bluetooth LE appearing in
2011-12, although the original (non-low energy) Bluetooth Classic technology is
older.  The use of the Bluetooth received signal strength reported by
smartphones to infer proximity has received attention since around 2014, mainly
in the context of studies on social interaction in indoor office settings and
only more recently making using of Bluetooth LE.    In summary, this work
highlights that (i) simple thresholding of received signal strength to estimate
proximity results in a high error-rate, (ii) by using standard machine learning
classifiers much better accuracy can be achieved, but since these are
supervised learning methods they require training data, which is generally
difficult and time-consuming to obtain (in these studies determining ground
truth involved manual observation of camera footage or shadowing of people by
observers).

One of the earliest studies is that reported in \cite{face2face2014}.  This makes use of Bluetooth (rather than Bluetooth LE).  Measurements are collected as two people carrying mobile handsets follow a prescribed path on a university campus, walking between two buildings and spending time indoors and outdoors.   The aim is to use received signal strength to detect when the two people are within 1.5m of each other.   It was found that simple thresholding of the received signal strength yields a high error rate of around 50\% even when people are in the proximity of one another for 10 mins.  Averaging of the measured signal strength values improves the accuracy somewhat, but since the signal strength was observed to vary significantly depending on whether people were indoors or outdoors then in order to achieve a low error rate the authors needed to employ multiple thresholds tuned to the particular environment used on the study, combined with use of a light sensor to detect operation indoors or out.   The latter meant that operation in evening and at night needed to be excluded.   A much larger follow-up study is also reported on in \cite{face2face2014} but since it lacks ground truth it is hard to draw quantitative conclusions regarding the accuracy with which proximity is estimated.

Around the same time the use of Bluetooth (again, not Bluetooth LE) for proximity detection was also considered by \cite{iccsac2015}.   In this study it is noted that phone orientation can have a substantial effect on received signal strength.  Water-filled cylinders with mobile handsets attached were used to collect baseline data on received signal strength that was then used to train a Decision Tree classifier.  In experiments involving 8 students interacting in an office environment, and with ground truth obtained via a human observer, proximity was estimated with an accuracy of around 80\%. 

More recently, \cite{percom2017} studies the use of Bluetooth LE for proximity sensing within 3m.   They use custom Bluetooth LE bracelets rather than smartphones and take measurements in an office-based workplace.  Ground truth is obtained by an observer logging all interactions, plus the office space is equipped with multiple static beacons whose transmissions are logged by the smartphones.   A Decision Tree classifier was trained using this data and an accuracy of around 80\% reported for Bluetooth LE settings similar to those used in Android.   Similarly,  last year \cite{sensys2019} reports on a study where people carry iBeacons.  Data is collected data for 24 people interacting in a 6m by 5m indoor space with ground truth on interactions being obtained via two video cameras covering the indoor space.   A Regression Tree classifier achieves an accuracy of around 80\%.

\subsection{iBeacons \& Indoor Localisation Using Bluetooth LE}
Apple introduced iBeacons using Bluetooth LE in mid-2013~\cite{ibeacon,ibeacon_wiki}.   These are typically placed in fixed locations indoors and transmit Bluetooth LE beacons.  These beacons typically transmit at low power so that they can only be detected when a receiver is relatively close and this allows them to be used by a mobile handset to roughly estimate its location within an office or shop.  Handsets use the received signal strength to estimate their distance from an iBeacon, but this is limited to whether the beacon is \emph{immediate}, \emph{near}, \emph{far} or has \emph{unknown} status.   Other manufacturers have since developed similar beacon technology.

There has, of course, been much interest in obtaining more accurate distance estimates so as to improve indoor localisation.   However, received signal strength measurements are known to exhibit large fluctuations, e.g. see~\cite{conext2017}, and so in more recent work it has been common to (i) try to combine received signal strength data with other measurements e.g. accelerometer and time of flight data, and (ii) employ machine learning methods to help map from received signal strength data to distance, similarly to the approaches adopted by the proximity detection community (see above).    See \cite{conext2017,conext2018,ipin2018}, and citations therein for recent work in this direction.


\section{Summary and Conclusions}
We report on measurements of the Bluetooth LE received signal strength taken on mobile handsets in a variety of common, real-world settings.  In summary, we find that the Bluetooth LE received signal strength can vary substantially depending on the relative orientation of handsets, on absorption by the human body, reflection/absorption of radio signals in buildings and trains.  Indeed we observe that the received signal strength need not decrease with increasing distance.  This suggests that the development of accurate methods for proximity detection based on Bluetooth LE received signal strength is likely to be challenging and time consuming.   Our measurements also suggest that it may be necessary for Bluetooth LE contact tracing apps to be combined with the adoption of new social protocols to yield benefits although this requires further study.  For example, placing phones on the table during meetings is likely to simplify proximity detection using received signal strength.  Similarly, carrying handbags with phones placed close to the outside surface.   In locations where the complexity of signal propagation makes proximity detection using received signal strength problematic, additional non-Bluetooth LE mechanisms may be required.

Looking ahead, further work is needed to try to quantify the error rates of proximity detection methods based on Bluetooth LE received signal strength.  In particular, it is important to distinguish between \emph{false positives} (where people are flagged as having been in contact with an infected person but in fact have not been) and \emph{false negatives} (where people are \emph{not} flagged as having been in contact with an infected person, but in fact were).    False positives are likely when the received signal strength does not decrease sufficiently quickly with distance, and false negatives when the received signal strength decreases too quickly.   Our data suggests that there may be significant potential for both types of error in common, real-world situations.   False positives are of concern since they mean that people may be led to unnecessarily self-isolate with associated disruption and perhaps also leading to a loss of confidence in the contact tracing app if the error rate is too high.  False negatives are of concern since they directly reduce the effectiveness of contact tracing for disease control, namely they mean that people in contact with an infected person may inadvertently spread the infection further.

\section*{Acknowledgements}
These measurements were taken in Dublin, Ireland during the Coronavirus lockdown.  Both authors would like thank their families for helping with the measurement experiments, they wouldn't have been possible without them.   The authors would also like to extend sincere thanks to Irish Rail for their generous permission to take measurements inside their train carriages.

\bibliographystyle{IEEEtran}
\bibliography{bibfile}

\end{document}